\documentclass[twocolumn,showpacs,amsmath,amssymb,pra,floatfix]{revtex4}
\pdfoutput=1
\usepackage{graphicx}
\usepackage{bm}
\usepackage{dsfont}
\usepackage{amsmath,amsfonts,amssymb}
\usepackage{epsfig}
\usepackage{psfrag}
\usepackage{pstricks}
\usepackage{float}
\newcommand{\bi}{\begin{itemize}}
\newcommand{\ei}{\end{itemize}\normalsize}

\newcommand{\lr}[1]{\left( #1 \right)}
\newcommand{\ket}[1]{|#1\rangle}
\newcommand{\bra}[1]{\langle #1|}

\newcommand{\mean}[1]{\langle #1 \rangle}

\newcommand{\vh}{\boldsymbol}
\newcommand{\no}{\nonumber}
\newcommand{\bea}{\begin{eqnarray}}
\newcommand{\eea}{\end{eqnarray}}
\newcommand{\be}{\begin{equation}}
\newcommand{\ee}{\end{equation}}

\newcommand{\f}[2]{\hat{\psi}_{#2}(\vh{#1})}
\newcommand{\fd}[2]{\hat{\psi}_{#2}^{\dag}(\vh{#1})}

\newcommand{\comm}[2]{\left[#1,#2\right]}
\begin{document}

\title{Non-equilibrium dynamics of bosonic atoms in optical lattices: \\Decoherence of many-body states due to spontaneous emission}

\author{H.~Pichler} \affiliation{Institute for Theoretical Physics,
  University of Innsbruck, A-6020 Innsbruck, Austria}\affiliation{Institute
  for Quantum Optics and Quantum Information of the Austrian Academy
  of Sciences, A-6020 Innsbruck, Austria}

\author{A.~J.~Daley} \affiliation{Institute for Theoretical Physics,
  University of Innsbruck, A-6020 Innsbruck, Austria}\affiliation{Institute
  for Quantum Optics and Quantum Information of the Austrian Academy
  of Sciences, A-6020 Innsbruck, Austria}

\author{P.~Zoller} \affiliation{Institute for Theoretical Physics,
  University of Innsbruck, A-6020 Innsbruck, Austria}\affiliation{Institute
  for Quantum Optics and Quantum Information of the Austrian Academy
  of Sciences, A-6020 Innsbruck, Austria}

\date{\today}

\begin{abstract}\label{txt:abstract}
We analyze in detail the heating of bosonic atoms in an optical lattice due to incoherent scattering of light from the lasers forming the lattice. Because atoms scattered into higher bands do not thermalize on the timescale of typical experiments, this process cannot be described by the total energy increase in the system alone (which is determined by single-particle effects). The heating instead involves an important interplay between the atomic physics of the heating process and the many-body physics of the state. We characterize the effects on many-body states for various system parameters, where we observe important differences in the heating for strongly and weakly interacting regimes, as well as a strong dependence on the sign of the laser detuning from the excited atomic state. We compute heating rates and changes to characteristic correlation functions based both on perturbation theory calculations, and a time-dependent calculation of the dissipative many-body dynamics. The latter is made possible for 1D systems by combining time-dependent density matrix renormalization group (t-DMRG) methods with quantum trajectory techniques.
\end{abstract}
\maketitle

\section{Introduction}

Recent experimental advances with ultracold quantum gases in optical lattices \cite{Jak98,Blo08,Lew07} have opened opportunities to explore novel phenomena in many-body lattice physics \cite{Sch08,Joe08,Bak09,She10,Gem09,Lin09,Win06,Hal09,Cat09,Sol10,Lig09,Cle09}, including aspects of quantum magnetism with bosonic and fermionic atoms, and possibilities for characterizing the phase diagram of the Fermi-Hubbard model. However, the production of strongly interacting many-body states at the very low temperatures required to reach many of these phases remains a key challenge in current experiments \cite{Tro09,Jor10,Wel09,Med10,McK10}.

In this context it is very important to be able to characterize and control heating processes arising in experiments. These can appear, e.g., via laser fluctuations that give rise to phase and amplitude noise on the lattice potential, through collisional losses of atoms, or via incoherent scattering of the lattice light. While heating of single atoms in dipole traps due to incoherent scattering was characterized a long time ago \cite{Gor80,Dal85}, and discussed recently in the specialized case of an optical lattice potential \cite{Ger10}, heating of many-body states in strongly interacting systems presents a new problem due to the interplay between essentially single- or few-body heating processes, and the characteristics of the many-body state. As a result of this interplay, different many-body states can be more or less sensitive to particular heating processes.

Here we analyze in detail the heating of bosonic atoms in an optical lattice due to incoherent scattering of light from the lasers forming the lattice. We choose this case both because it provides a clear example in which to study the interplay between the atomic physics of the heating process and the many-body physics of the state, and because incoherent scattering is expected to be the dominant heating mechanism in recently analyzed experiments \cite{Tro09}. The Bose Hubbard model corresponding to bosonic atoms in the lowest-energy Bloch band of an optical lattice is defined via the Hamiltonian ($\hbar=1$)
\be \label{eq:HBH}
H_{BH}=-J\sum_{\langle i,j \rangle} b_i^{\dag}b_j +\frac{1}{2} U\sum_i {b_i^{\dag}}^2 {b_i}^2,
\ee
where the first term is a kinetic energy describing the hopping of bosons on the lattice with amplitude $J$ with $b_i$ ($b^\dagger_i$) bosonic destruction (creation) operators, and the second term is an onsite interaction with strength $U$. The phase diagram of the Bose Hubbard model contains a superfluid phase for $J\gg U$, and a Mott insulator phase for $J\ll U$, connected by a quantum phase transition. This phase diagram has been discussed extensively in the literature using both mean field and exact Quantum Monte Carlo techniques (see Ref. \cite{Tro09}, and references cited).  For reference below, we note that the critical point for the quantum phase transition appears at $(U/J)_c\approx 3.37$ in 1D \cite{Kue00}, whereas in 2D and 3D the quantum phase transition at unit filling is closer to the mean field value of $(U/(zJ))_c\approx 5.8$ with $z$ the number of nearest neighbors for each lattice site. Our goal below is to characterize the effects of incoherent scattering of lattice light on many-body states for various system parameters. The non-equilibrium dynamics of this process are described by a master equation, which includes the coherent atomic dynamics in form of a multiband Bose Hubbard model, as well as an incoherent part describing the effects of spontaneous emission. In this context we observe important differences in the heating for strongly and weakly interacting regimes, as well as a strong dependence on the sign of the laser detuning.

A key feature of the physics here is that it is not sufficient to determine the total rate of energy increase in the system in order to characterize the heating. This is because some single-particle excitations do not thermalize on typical experimental timescales (e.g., individual atoms excited to higher Bloch bands). Indeed, the mean rate of energy increase is independent of the interactions and of the sign of the laser detuning, as is found for a single atom (see, e.g., Ref.~\cite{Ger10}). Instead, the change in the many-body state must be characterized in terms of characteristic correlation functions for the state, e.g., the single particle density matrix, which characterizes off-diagonal order in the superfluid regime.

We compute the time dependence of characteristic correlation functions based both on perturbation theory calculations, and a time-dependent calculation of the dissipative many-body dynamics. The time dependent dynamics are described by a many-body master equation, and can be computed in the mean-field limit using a density-matrix Gutzwiller approach. For 1D systems they can be computed exactly by combining time-dependent density matrix renormalization group (t-DMRG) methods with quantum trajectory techniques \cite{Dal09}. We show that in the weakly interacting regime, bosons are strongly susceptible to heating in the sense that long-range order in the superfluid ground state is destroyed by a localization mechanism in spontaneous emission events. In contrast, a Mott Insulator ground state, in which each atoms is already exponentially localized at a particular lattice site, is very robust against spontaneous emissions. The rate of destruction of long-range order depends on the total scattering rate, not on the energy input into the system, and so is much more rapid for red-detuned lattices than for blue-detuned lattices.

This article is organized as follows: in Sec.~\ref{sec:model} we briefly review the description of a single atom in an optical lattice including spontaneous emission and then present the model we use to describe the situation of many atoms. In Sec.~\ref{sec:quantification} we present key quantities characterizing spontaneous emission such as the scattering rate, the total increase in energy and the effect on the key correlation functions for a system in the ground state of a the Bose Hubbard model. In Sec.~\ref{sec:QT_TEBD} and \ref{sec:Gutzi} we present the results of fully time dependent calculations based on an exact numerical calculation for 1D lattices, and a Gutzwiller mean field approach for 3D lattices, respectively. In Sec.~\ref{sec:conclusion} we then present a summary and outlook.

\section{Model}\label{sec:model}

\label{sec:model} We consider bosonic atoms in an optical lattice,
which is generated by a far-detuned laser fields via the AC-Stark
shift, and study how spontaneous emission affects the many-body state
of the atoms. We will describe atomic dynamics in terms of master
equations,\[
\dot{\rho}=-i\left[H,\rho\right]+{\cal L}\rho,\]
with $\rho$ the reduced density operator of the atoms, where we trace
over the bath of vacuum radiation modes, $H$ is a multiband Hubbard Hamiltonian
for bosons in an optical lattice, and ${\cal L}$ is a Liouvillian representing
the effects of spontaneous emission.

In Sec.~\ref{sec:singleparticle} we will first write out this equation for two-level
systems with the off-resonant excited state eliminated.  Such master equations have been developed in the context
of laser cooling \cite{Metcalf_book,Minogin_book}, mainly in a single particle context \cite{Dal85,Mar93,Ger10}, and we will adapt and
generalize them to the present problem. We will also discuss the differences in the scattering rates
and similarities in the heating rates for blue and red detuned laser
light in this case.  We note that for very large detunings the assumptions of a two-level atom and
rotating wave approximation (RWA) break down, but we stick to this model because
the two-level results are compact and transparent, and are readily generalized
to include more levels, and contributions from counterrotating terms.
Finally, in Sec.~\ref{sec:nparticle} we generalize to $N$  atoms, including
a discussion of interactions and non-idealities in typical experimental
setups.

\subsection{Single particle case}
\label{sec:singleparticle}

In this subsection, we briefly review the dynamics of a single two-level atom with mass $m$ and internal states $\ket{g}$ and $\ket{e}$ in an optical field (for a more detailed discussion of these dynamics see
Ref.~\cite{Ger10}). Below we begin from the optical Bloch equations including the atomic motion, and then derive an effective master equation for the ground state $\ket{g}$ in the limit of large laser detuning. We then derive a form for the master equation expanded in terms of Wannier functions for the lattice potential, before discussing the key features of heating in a deep lattice, and how this heating depends on the sign of the laser detuning.

\subsubsection{Optical Bloch Equations with motion}

The motion of a two-level atom driven by a laser field and undergoing spontaneous emission is described by Optical Bloch Equations ($\hbar=1$)\cite{Dal85,Mar93},
\bea\label{eq:OBE}
\dot{\rho}&=&-i\left[H,\rho\right]+\\
&+&\Gamma\int d^2\vh{u}N(\vh{u})\left( C_{\vh{u}}\rho\,C^{\dag}_{\vh{u}}-\frac{1}{2}C^{\dag}_{\vh{u}}C_{\vh{u}}\rho-
\frac{1}{2}\rho\,C^{\dag}_{\vh{u}}C_{\vh{u}}\right).\nonumber
\eea
with atomic Hamiltonian
\begin{equation}\label{eq:single_article_Hamiltonian}
H=\frac{\hat{\vh{p}}^{\,2}}{2m}-\Delta \ket{e}\bra{e}-\left(
\ket{g}\bra{e}\frac{\Omega(\hat{\vh{x}})}{2}
+{\rm h.c.}\right).
\end{equation}
Here $\rho$ is the reduced density matrix of the two-level atom  $\{\ket{g},\ket{e}\}$. In Eq.~(\ref{eq:OBE}) $\Gamma$ is the decay rate for the excited state $e$. The Lindblad operator $C_{\vh{u}}=\ket{g}\bra{e}e^{-ik_{eg}\vh{u}\cdot
\hat{\vh{x}}}$ describes the return of the atomic electron $\ket{e}\rightarrow\ket{g}$ to the ground state after a photon emission in direction $\vh{u}$, including the associated recoil kick to the atom. Here $k_{eg}\equiv\omega_{eg}/c\approx k_L$ is the wavenumber associated with the atomic transition frequency $\omega_{eg}$, and $N(\vh{u})$ is the distribution of directions for the emitted photons. If we denote the unit vector along the dipole moment of the transition with $\hat{\vh{d}}$, this distribution is given by
\be
N(\vh{u})=\frac{3}{8\pi}\left(1-\left(\vh{u}\cdot\hat{\vh{d}}\right)^2\right).
\ee
The atomic Hamiltonian (\ref{eq:single_article_Hamiltonian}) contains the kinetic energy and the laser interactions. The laser driving the atom has optical frequency $\omega_L\equiv ck_L$ with detuning $\Delta =\omega_L - \omega_{eg}$ from atomic resonance. The speed of light is denoted by $c$.  The laser interaction is characterized by a spatially dependent Rabi frequency $\Omega(\vh{x})$ proportional to the electric field of the laser and the atomic dipole moment $d$.
Eqs.~(\ref{eq:OBE}) and (\ref{eq:single_article_Hamiltonian}) are written in a frame rotating with the laser frequency, i.e. the optical frequencies have been eliminated.

\subsubsection{Elimination of the excited state}

In the limit of small saturation and large detuning $|\Delta|\gg\Omega,\Gamma$ we can eliminate the excited state adiabatically to obtain a master equation for the external degrees of freedom only. Denoting the density operator for the motion again with $\rho$ this reads:
\begin{align}\label{eq:singl_part_ad_el_master_eq}
\frac{d}{dt}\rho&=-i(H_{\textrm{eff}} \rho-\rho  H_{\textrm{eff}}^{\dag})+{\mathcal{J}}{\rho},
\end{align}
where the non-hermitian effective Hamiltonian is given by
\bea H_{\textrm{eff}}&=&\frac{\hat{\vh{p}}^{\,2}}{2m}+\frac{|\Omega(\hat{\vh{x}})|^2}{4\Delta}-i\frac{1}{2}\frac{\Gamma|\Omega(\hat{\vh{x}})|^2}{4\Delta^2}\no\\
&\equiv&\frac{\hat{\vh{p}}^{\,2}}{2m}+V_{\rm opt}(\hat{ \vh{x}}) -i\frac{\gamma(\vh{x})}{2}.
\eea
Here we have identified the spatially dependent AC Stark shift with the optical potential $V_{\rm opt}(\vh{x})$
and $\gamma(\vh{x})$ with the rate of light scattering. The recycling term in Eq.~(\ref{eq:singl_part_ad_el_master_eq}) is given by
\be\label{eq:single_particle_recycling}{\mathcal{J}}{\rho}=\Gamma\int d^2\vh{u}\,N(\vh{u})\left[
e^{-ik_{eg}\vh{u}\cdot\hat{\vh{x}}}\frac{\Omega(\hat{\vh{x}})}{2\Delta}\right]\rho \left[ e^{ik_{eg}\vh{u}\cdot\hat{\vh{x}}}\frac{\Omega^{\ast}(\hat{\vh{x}})}{2\Delta}\right],
\ee
where the operators $c_{\vh{u}}(\hat{\vh{x}})\equiv e^{-ik_{eg}\vh{u}\cdot\hat{\vh{x}}}\Omega(\hat{\vh{x}})/(2\Delta)$ correspond to absorption of a laser photon followed by the scattering of a spontaneous photon in the direction $\vh{u}$.
We will write the many-body master equation in a similar form in Sec.~\ref{sec:nparticle}.

\subsubsection{Expansion in Wannier modes}

For a periodic optical potential we can expand the master equation in a basis of real Wannier functions that are exponentially localized at each lattice site \cite{Koh59} (as is done in the standard derivation of the Bose-Hubbard model \cite{Jak98}).  Here we denote the Wannier function at lattice site $\vh{i}$ in the Bloch band $\vh{n}$ as $w_{\vh{i}}^{(\vh{n})}(\vh{x})$. Anticipating the  $N$ bosons case below, we introduce second quantized mode operators $b_{\vh{i}}^{(\vh{n})}$ that annihilate a particle at a site $\vh{i}$ in the band $\vh{n}$, and obey the usual bosonic commutation relations.

In the tight binding approximation, valid for a sufficiently deep lattice, we then obtain for the effective Hamiltonian
\begin{align}\label{eq:single_particle_wannier_hamiltonian}
 H_{\textrm{eff}}&=-\sum_{\vh{n},\langle \vh{i},\vh{j}\rangle}J_{\vh{i},\vh{j}}^{(\vh{n})}b_{\vh{i}}^{(\vh{n})\,\dag}b_{\vh{j}}^{(\vh{n})}+
\sum_{\vh{n},\vh{i}}\varepsilon^{(\vh{n})}b_{\vh{i}}^{(\vh{n})\,\dag}b_{\vh{i}}^{(\vh{n})}+\no\\
&\quad-\frac{i}{2}\sum_{\vh{n},\vh{m},\vh{i}}\gamma^{(\vh{n},\vh{m})}b_{\vh{i}}^{(\vh{n})\,\dag}b_{\vh{i}}^{(\vh{m})}.
\end{align}
The first line in Eq.~(\ref{eq:single_particle_wannier_hamiltonian}) is a multiband Hubbard Hamiltonian, and the second line are decay terms. The relevant matrix elements are:
\begin{align}
 J_{\vh{i},\vh{j}}^{(\vh{n})}&=-\int d^3x\, w_{\vh{i}}^{(\vh{n})}(\vh{x}) \left(\frac{\hat{\mathbf{p}}^2}{2m}+V_{\rm opt}(\vh{x}) \right)w_{\vh{j}}^{(\vh{n})}(\vh{x})\\
 \varepsilon^{(\vh{n})}&=\int d^3x\, w_{\vh{i}}^{(\vh{n})}(\vh{x}) \left( \frac{\hat{\mathbf{p}}^2}{2m}+V_{\rm opt}(\vh{x}) \right)w_{\vh{i}}^{(\vh{n})}(\vh{x})\\
 \gamma^{(\vh{n},\vh{m})}&=\int d^3x\, w_{\vh{i}}^{(\vh{n})}(\vh{x}) \gamma(\vh{x}) w_{\vh{i}}^{(\vh{m})}(\vh{x}).\label{eq:single_paericle_gamma_wannier}
\end{align}
In practice the hopping rates $J_{\vh{i},\vh{j}}^{(\vh{n})}$ can easily calculated from the band structure.
For the Lindblad operators in the recycling term we obtain
\be\label{eq:single_particle_wannier_jump}
 c_{\vh{u}}=\sum_{\vh{n},\vh{m},\vh{i}}\int d^3x w_{\vh{i}}^{(\vh{n})}(\vh{x})e^{-ik_{eg}\vh{u}\cdot{\vh{x}}}\frac{\Omega(\vh{x})}{2\Delta} w_{\vh{i}}^{(\vh{m})}(\vh{x})b_{\vh{i}}^{(\vh{n})\,\dag}b_{\vh{i}}^{(\vh{m})}.\ee
They describe the redistribution of the atoms in the Bloch bands due to absorption of a laser photon from the optical lattice lasers followed by emission of a photon. We note that because the Wannier functions are exponentially localized, spontaneous emission processes coupling atoms between neighboring sites are small, and these matrix elements have been neglected above.

In Sec.~\ref{sec:nparticle} we will generalize these equations to include inter-particle interactions.

\subsubsection{Heating processes for red and blue detuned light}
\label{sec:heatredblue}

We now consider the differences in the description of the heating processes for red-detuned ($\Delta<0$) and blue-detuned ($\Delta>0$) light. These cases are distinguished by the sign of $V_{\rm opt}(\vh{x})$, which results for the blue-detuned case in the minima of the potential occurring at the minima of the field intensity, i.e., the minima of $\Omega(\vh{x})$, but for the minima to occur at the maximum field intensity in the red-detuned case.

In this subsection we will write the potential along 1D for simplicity, but these concepts are readily generalized to the 3D case. In order to make a comparison where a minimum of the potential is always centered at $x=0$, we write the Rabi frequency to behave as $\Omega(x)= \Omega_0\cos(k_L x)$ for a red-detuned lattice, and $\Omega(x)=\Omega_0\sin(k_Lx)$ for a blue-detuned lattice. The optical potential then behaves as $V_{\rm opt}(x)=(|\Omega_0|^2/|4\Delta|) \sin^2(k_Lx)$ for the blue-detuned case, and $V_{\rm opt}(x)=-(|\Omega_0|^2/|4\Delta|) \cos^2(k_Lx)=(|\Omega_0|^2/|4\Delta|) \sin^2(k_Lx)-(|\Omega_0|^2/|4\Delta|)$ in the red-detuned case. In both cases, the minima of the potential occur at $x=0,\pm\pi/k_L,\pm2\pi/k_L\dots\equiv0,\pm a,\pm 2 a\dots$. The different form of $\Omega(x)$ for red- and blue-detuned lattices has a large effect on the type of heating processes that are possible, as we can see by considering the behavior of the field around the minima of the field where the atoms are trapped, where for red-detuned light ($\Omega_{\rm red}(x)\approx\Omega_0(1-(k_Lx)^2/2)$), and for blue-detuned light ($\Omega_{\rm blue}(x)\approx\Omega_0 k_Lx$).

\begin{figure}[tb]\centering
\includegraphics[width=8cm]{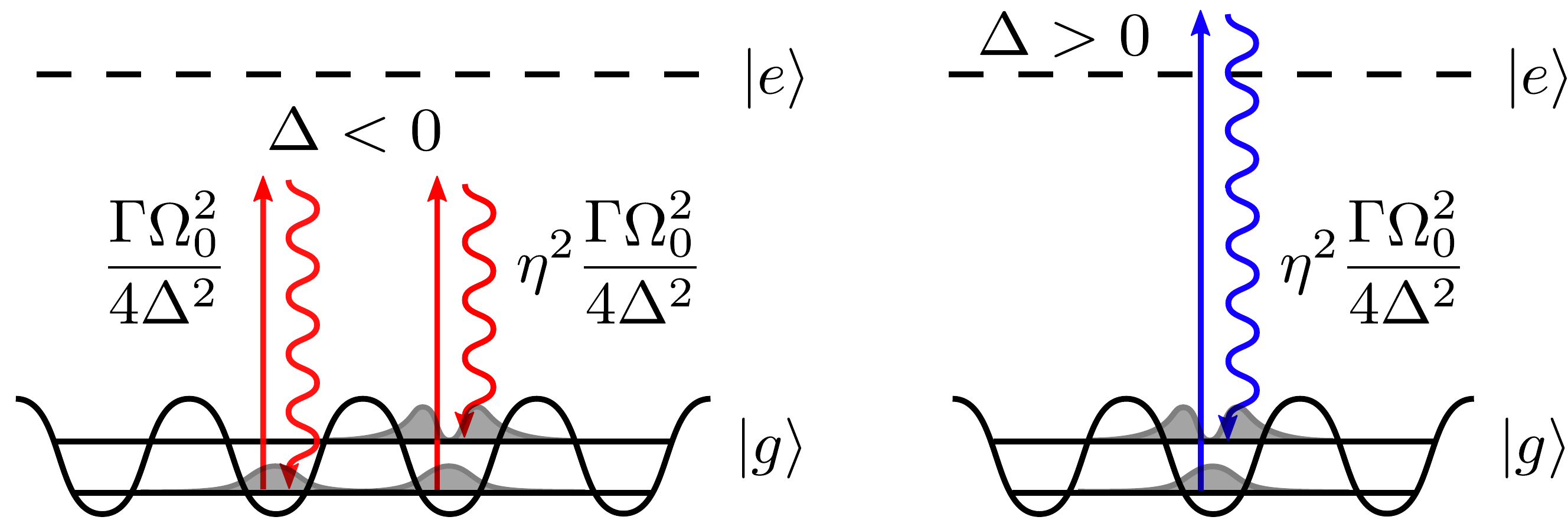}
\caption{Schematic picture of the leading-order heating processes
via scattering of laser photons in a red-detuned (left) and a blue-detuned (right) lattice, in the \emph{Lamb-Dicke} limit (see text).
In a red-detuned lattice the fastest process
returns the scattered atom to the lowest band, whilst this process is strongly suppressed
in the blue detuned lattice. In both cases there are higher
order processes in the Lamb-Dicke parameter $\eta$ of the
lattice, in which the atoms are scattered into higher
Bloch bands. In the limit of a deep lattice, these processes occur at identical rates in the red-detuned and blue-detuned cases.}\label{fig:scattering_rate_sketch}
\end{figure}

To obtain a simple picture for the scattering processes, we can consider atoms that are tightly trapped in the lattice, so that at each site and for each dimension, the \emph{Lamb Dicke parameter}, $\eta=k_La_0$, which compares the extension $a_0$ of the lowest band Wannier function to the wavelength of scattered photons ($2\pi/k_L$) is a small parameter. The dependence of this parameter on the depth of the lattice $V=|\Omega_0|^2/(4|\Delta|)$, is given by $\eta=k_La_0=(4V/E_R)^{-1/4}$ (where $E_R=k_{L}^2/(2m)$ is the recoil energy).  In the limit of deep lattices ($V\gg E_R$), the Wannier functions can be approximated with harmonic oscillator wave functions around the potential minima and the integrals in Eq.~(\ref{eq:single_paericle_gamma_wannier}) and (\ref{eq:single_particle_wannier_jump}) may be evaluated in a Lamb Dicke limit ($\eta\ll 1$), expanding the Rabi frequency and the plane wave in a power series in $k_Lx$. The leading order scattering processes in this limit are depicted in Fig.~\ref{fig:scattering_rate_sketch}. We then obtain that the scattering rate for particles initially trapped in the lowest band to lowest order in the Lamb Dicke parameter is given by
\bea
\Gamma_{\rm scatt}=\gamma^{(\vh{0},\vh{0})}&\rightarrow\left\{
                  \begin{array}{ll}
                    \frac{\Gamma|\Omega_0|^2}{4\Delta^2}, & \hbox{$\Delta<0$;} \\
                    \frac{\Gamma|\Omega_0|^2}{4\Delta^2}\eta^2, & \hbox{$\Delta>0$.}
                  \end{array}
                \right.
\eea
As is expected from the fact that atoms in the blue-detuned case are at the minimum of the light intensity, the scattering rate is reduced substantially in this case, by a factor of $\eta^2$.
This corresponds to a substantial decrease in the probability of atoms being scattered back to the lowest Bloch band in this case, where in the Harmonic oscillator approximation the rate of scattering returning atoms to the lowest band is given by
\bea  \Gamma_{\textrm{scatt}}^{\textnormal{\tiny{l.b.}}}\rightarrow \left\{
                                                                                                     \begin{array}{ll}
                                                                                                       \frac{\Gamma|\Omega_0|^2}{4\Delta^2}, & \hbox{$\Delta<0$;} \\
                                                                                                       \frac{\Gamma|\Omega_0|^2}{4\Delta^2}\eta^4, & \hbox{$\Delta>0$.}
                                                                                                     \end{array}
                                                                                                   \right.\no\eea
Whereas scattering into the lowest band is the dominant term (order $\eta^0$) in the red-detuned lattice, the leading order term for the blue-detuned lattice is scattering into the first excited band, which is of order $\eta^2$. In the Harmonic oscillator approximation this coupling to the first excited band occurs at the identical rate in the red-detuned lattice .

While the red-detuned and blue-detuned cases exhibit considerably different scattering rates, the rate at which the energy of the atom $E=\langle  H \rangle$ increases is identical for red and blue detuning, and completely independent of the initial motional state of the atom (see Appendix \ref{ap:heting_rate}),
\begin{equation}
\dot E=\frac{\Gamma|\Omega_0|^2}{4\Delta^2}E_R.
\end{equation}
In the harmonic oscillator approximation, the independence of this rate on the sign of the detuning can be seen clearly, because an increase in energy arises solely from coupling to higher bands, and whilst the rate of scattering events returning particles to the lowest band are very different for blue-detuned and red-detuned light, the rate with which particles are scattered to higher bands is identical in the two cases.
When the same computation is performed with Wannier functions, the lowest Bloch band has a finite energy width, and thus a small energy increase is obtained from atoms scattered into the lowest band. The heating rate in the red- and blue- detuned cases remains identical, however, with the scattering rate into higher bands in the red-detuned case being slightly reduced to compensate the energy increase from scattering into the lowest band.

We note at this point that the rate of energy increase in the system does not completely characterize the heating process, and we will show below that for many-body systems, heating in blue- and red- detuned lattices can function very differently, despite the fact that the rate of energy increase remains independent of the sign of the detuning. A key physical characteristic of spontaneous emission processes when characterizing how the many-body state changes is that they tend to localize particles undergoing the scattering event on a lengthscale of the wavelength of the emitted photon. This will be described in more detail in Sec.~\ref{sec:localisation}.

\subsubsection{Remarks on the master equation}

We conclude with a few remarks regarding the validity of the above model. The assumption of a two-level system here is clearly an oversimplification in some respects. However, it is very clear how to generalize these results to more realistic atomic models and experimental setups.

First, in order to generate an isotropic 3D cubic optical lattice, it is typical to use three laser beams that are independent, either because they are slightly detuned or because they have orthogonal polarizations. The optical potential is a sum of optical potentials for all three beams, and all terms in the master equation must be summed over contributions from the different beams. In the calculations below we explicitly choose different lattice depths in different directions, and perform this summation, in order to produce results that are as close as possible to current experiments. The sum of the effective scattering rate $|\Omega_0|^2 \Gamma / (4\Delta^2)$ over the beams will be denoted $\gamma_0$.

Second, for typical parameters, the lattice lasers are so far detuned that the rotating wave approximation is not strictly valid. However, in the limit where the excited state can be eliminated, this does not change the effective description of the physics except for small quantitative modifications in the prefactor of the scattering rate and the optical potential.

Finally, in multi-level atoms, the far-detuned lasers will couple to many excited states. Again, in the limit of large detuning, all of these states can be adiabatically eliminated to produce the same effective model with prefactors that arise from correctly summing the contributions from all excited levels.

Similar remarks also apply to the $N$-Atom Master equation that we present below.

\subsection{$N$-Atom Case}
We now present the full model for $N$ atoms in the lattice. Below we first state the full master equation assuming two-level atoms, and discuss the origins of each term, including terms arising from interactions and collisional loss in the system.
\label{sec:N_atom_Model}
\label{sec:nparticle}

\subsubsection{Master equation for $N$ atoms}

For $N$ bosonic atoms the evolution of the reduced system density operator $\rho$ is given by a master equation which we write in using second quantization. Eliminating the atoms in the excited state and defining  field operators $\hat{\psi}(\vh{x})$ with bosonic commutation relations $\left[ \hat{\psi}(\vh{x}), \hat{\psi}^\dagger(\vh{y})\right]=\delta(\vh{x}-\vh{y})$ for atoms in the ground state, we derive again a master equation of the form
\bea\label{eq:master_equation}
\dot \rho&=&-i \left( H_{\textrm{eff}}\rho-\rho H_{\textrm{eff}}^\dag \right) +\mathcal{J}\rho,
\eea
with non-hermitian effective Hamiltonian
\bea
H_{\rm eff} &=&H_0 + H_{\rm eff}^{\rm rad}+ H_{\rm eff}^{\rm coll}.
\eea
The first contribution to the effective Hamiltonian, $H_0$, is the term describing motion of single atoms in the optical lattice,
\begin{flalign}
H_0=\int d^3x \hat{\psi}^{\dag}(\vh{x})
\left(
-\frac{\nabla^2} {2m}+V_{\rm opt}(\vh{x})
\right)\hat{\psi}(\vh{x}),&&\label{eq:H1}
\end{flalign}
which is the same form derived in Sec.~\ref{sec:singleparticle}.

The radiative part of the master equation describing the couplings of the atoms to the vacuum modes of the electromagnetic field are contained in the effective Hamiltonian
\begin{widetext}
\begin{flalign}
H_{\rm eff}^{\rm rad}&= \iint d^3xd^3y \frac{\Gamma\Omega(\vh{y})\Omega^{\ast}(\vh{x})}{4\Delta^2}G(k_{eg}(\vh{x}-\vh{y})) \fd{x}{}\fd{y}{} \f{y}{}\f{x}{} &&\label{eq:H2}\\
&\quad-i\frac{1}{2}\int d^3x \frac{\Gamma| \Omega(\vh{x})|^2}{4\Delta^2} \hat{\psi}^{\dag}(\vh{x})\hat{\psi}(\vh{x})
-i\frac{1}{2}\iint d^3xd^3y \frac{\Gamma\Omega(\vh{y})\Omega^{\ast}(\vh{x})}{4\Delta^2}F(k_{eg}(\vh{x}-\vh{y})) \fd{x}{}\fd{y}{}\f{y}{}\f{x}{},\label{eq:H3}
\end{flalign}
and recycling term
\begin{flalign}
\mathcal{J}\rho&=\iint d^3x d^3y \frac{\Gamma\Omega(\vh{x})\Omega(\vh{y})}{4\Delta^2} F(k_{eg}(\vh{x}-\vh{y}))
 \hat{\psi}^{\dag}(\vh{x})\hat{\psi}(\vh{x})\rho
\hat{\psi}^{\dag}(\vh{y})\hat{\psi}(\vh{y}),&&\label{eq:dissipation}
\end{flalign}
with functions $F$ and $G$ defined as
\begin{flalign}
F(\vh{\xi})&=\int d^2\vh{u}\,N(\vh{u})e^{-i\vh{u}\cdot\vh{\xi}}&&\no\\
&=\frac{3}{2}\left\{\frac{\sin\xi}{\xi}\left(1-(\hat{\vh{d}}\cdot\hat{\vh{\xi}})^2\right)+\right.
\left.\left(1-3(\hat{\vh{d}}\cdot\hat{\vh{\xi}})^2\right)\left(\frac{\cos\xi}{\xi^2}-\frac{\sin\xi}{\xi^3}\right)\right\},\label{eq:F_def}\\
G(\vh{\xi})&=-\frac{1}{\xi^3}\mathcal{P}\int_{-\infty}^{\infty}\frac{d\zeta}{2\pi}\frac{\zeta^3}{\zeta-\xi}F(\zeta\vh{\xi}/\xi)&&\no\\
&=\frac{3}{4}\left\{-\left(1-(\hat{\vh{d}}\cdot\hat{\vh{\xi}})^2\right)\frac{\cos\xi}{\xi}+\right.
\left.\left(1-3(\hat{\vh{d}}\cdot\hat{\vh{\xi}})^2\right)\left(\frac{\sin\xi}{\xi^2}+\frac{\cos\xi}{\xi^3}\right)\right\},\label{eq:G_def}
\end{flalign}
\end{widetext}
where $\mathcal{P}$ denotes the principal value integral.

Finally, we have an effective Hamiltonian accounting for short range collision physics in the presence of laser fields, as well as associated losses
\begin{flalign}
H_{\rm eff}^{\rm coll}=\int d^3x \left( g(\vh{x})-i\frac{1}{2}\gamma_2(\vh{x})\right) \hat{\psi}^{\dag}(\vh{x})\hat{\psi}^{\dag}(\vh{x})\hat{\psi}(\vh{x})\hat{\psi}(\vh{x}),&&\label{eq:H4}
\end{flalign}
where the functions $g$ and $\gamma_2$ will be discussed below.

We will now give a more detailed discussion of the radiative and collisional contributions to the master equation, as well as commenting on the validity of these equations in each case.

\emph{Radiative terms} -- As mentioned above, radiative processes are contained in the effective Hamiltonian $H_{\rm eff}^{\rm rad}$ and the recycling term $\mathcal{J}$. These contributions to the master equation are obtained by eliminating the vacuum modes of the radiation field for an ensemble of two-level atoms, as discussed by Lehmberg \cite{Leh70,Leh70_2}, followed by an adiabatic elimination of the excited state. The present master equation is a straightforward generalization of Lehmberg's $N$-atom Bloch equations by including the quantized motion of the atoms, and by writing the master equation in second quantized form. The first line in $H_{\rm eff}^{\rm rad}$ is the dipole-dipole interaction due to exchange of photons between the atoms. The second line contains a single particle decay term corresponding to the absorption of a laser photon, followed by spontaneous emission, as discussed in Sec.~\ref{sec:singleparticle}. The second term in Eq.~(\ref{eq:H3}) is a collective radiative term associated with super- and subradiance. The functions $F$  and $G$ appearing in Eqs.~(\ref{eq:H2},\ref{eq:H3}) and (\ref{eq:dissipation}) are defined in Eqs.~(\ref{eq:F_def},\ref{eq:G_def}). For distances much smaller than the optical wavelength $\xi=k_{eg}|\mathbf{x}-\mathbf{y}| \ll 1$, i.e. for particles on the same lattice site, the function $G$ approaches the static dipole-dipole interaction diverging as $r^{-3}$, while $F(0)=1$. For distances larger than the wavelength, i.e. particles on distant lattice sites, both $G$ and $F$ fall off in an oscillatory manner on a lengthscale set by the wavelength of the emitted photons. A plot of $F(\xi)$ can be found in Fig.~\ref{fig:F_plot}.

We conclude the discussion of the radiative terms with three remarks. First, the recycling term $\mathcal{J}$ in Eq.~($\ref{eq:dissipation}$) involves Lindblad operators in the form of atomic densities $\hat{\psi}^{\dag}(\vh{x})\hat{\psi}(\vh{x})$ smeared out by the function $F(k_{eg}(\vh{x}-\vh{y}))$. Thus a spontaneous emission event will localize a particle within a wavelength. This is a key mechanism behind destruction of long-range order due to spontaneous emission processes, and we will return to this discussion below in Sec.~\ref{sec:localisation} and Sec.~\ref{sec:quantspdm}.
Second, it is easily checked that the effective Hamiltonian $H_{\rm eff}^{\rm rad}$ together with the recycling term ($\ref{eq:dissipation}$) give a trace preserving master equation in the usual Lindblad form. This follows from the commutation relations of $\hat{\psi}(\vh{x})$ together with $F(0)=1$.
Thirdly, we note that the dipolar function $G(\vh{\xi})$ must be understood as regularized on a short distance scale of molecular interactions where the two-level model underlying the derivation of the radiative master equation breaks down. In typical experimental situations, where the lattice is very far detuned, the dipole-dipole interaction term will give only very small corrections to the effects of normal collisional interactions, which we discuss in the next paragraph. In the derivation of a Hubbard model, these dipole-dipole interaction terms will give small corrections to the onsite interaction energy, and will be absorbed into this value for the purposes of the discussions in following sections.

\emph{Collisional terms} -- We now turn to the collisional terms describing short range scattering.  In the absence of laser light, and for scattering at sufficiently low energies and densities, this short-range physics provides a boundary condition for the two-body scattering wavefunctions at longer distances, so that the resulting dynamics can be described as an effective two-body interaction with a single parameter, the scattering length $a_s$ \cite{Stringari_book,Pethick_book,Chi10}. Thus, this short range collision physics is accounted for by a contact potential in the many particle Hamiltonian, as represented by $H_{\rm eff}^{\rm coll}$ in Eq.~(\ref{eq:H4}) with
 $g=4\pi \hbar^2 a_s/m$. The scattering length $a_s$ can be calculated by solving a set of coupled channel equations for scattering of atoms \cite{Gao05}. It is, at least on a conceptual level, straightforward to include in these channels not only collisional processes but also light assisted collisional interactions. An example is provided by the discussion of the optical Feshbach resonances by Fedichev et al. \cite{Fed96,The04,Zel06}, where for red detuned laser light excited electronic states of the atom-atom complex provide resonances and thus a resonant enhancement of the ground state scattering length. Such a light-modified scattering length will reflect the local laser intensity, and thus, in principle, be spatially dependent. We account for this by writing an intensity-dependent and spatially varying contact coupling $g(\mathbf{x})$ in Eq.~(\ref{eq:H4}). Away from the optical Feshbach resonance we expect $g(\mathbf{x})\sim 4\pi \hbar^2 a_s/m$. More important, radiative loss in these processes can be accounted for by an imaginary part of the a scattering length, which will again be intensity and thus spatially dependent, as discussed in the context of optical Feshbach resonances. We account for this loss by including an intensity dependent imaginary contribution to the scattering length $-i\gamma_2$ in the contact interaction (\ref{eq:H4}). Along a similar reasoning losses in collisions between chemically reactive molecules have recently been modeled by imaginary scattering length.

In general, there can be corrections to collisional interactions due to higher order processes, e.g., three-body collisions that can generate three-body losses. These are neglected here. Also, in the discussions below we would like to separate the role of spontaneous emission events from that of light-assisted collisions. As a result, we will set $\gamma_2=0$ for the purposes of calculations exploring the effects of heating due to spontaneous emissions.

\subsubsection{The master equation in a Wannier basis}

\emph{Multi-band Hubbard model} --
When the potential corresponds to an optical lattice, we obtain a multi-band Bose-Hubbard model from Eq.~(\ref{eq:master_equation}) for the coherent part of the evolution by expanding the field operators in a Wannier basis, $\psi(\vh{x})=\sum_{\vh{n,i}}w_{\vh{i}}^{(\vh{n})}(\vh{x})b_{\vh{i}}^{(\vh{n})}$, and applying the assumption of local tunneling and interactions in a deep lattice \cite{Jak98}, e.g. for an isotropic 3D lattice,
\begin{widetext}
\begin{align}
H&=-\sum_{\vh{n},\langle\vh{i},\vh{j}\rangle}J_{\vh{i},\vh{j}}^{(\vh{n})}b_{\vh{i}}^{(\vh{n})\,\dag}b_{\vh{j}}^{(\vh{n})} + \sum_{\vh{n},\vh{i}}\varepsilon^{(\vh{n})}b_{\vh{i}}^{(\vh{n})\,\dag}b_{\vh{i}}^{(\vh{n})}
+\sum_{\vh{i},\vh{k},\vh{l}\vh{m},\vh{n}}\frac{1}{2}U^{(\vh{k,l,m,n})}b_{\vh{i}}^{(\vh{k})\,\dag}b_{\vh{i}}^{(\vh{l})\,\dag}b_{\vh{i}}^{(\vh{m})}b_{\vh{i}}^{(\vh{n})},
\label{eq:multibandham}
\end{align}
with tunneling rates $J_{\vh{i},\vh{j}}^{(\vh{n})}$ and onsite interaction energy shifts $U^{(\vh{k,l,m,n})}$ arising from the contact and dipole-dipole interactions in Eqs.~(\ref{eq:H2},\ref{eq:H4}). Below we will denote $J_{\vh{i},\vh{j}}^{(\vh{n})}$ and $U^{(\vh{n,n,n,n})}$ for the lowest band as $J$ and $U$, respectively.
The remaining part of the master equation in this basis takes on a similar form to that in Sec.~\ref{sec:singleparticle}:
\bea
\dot\rho&=&-i\comm{H}{\rho}+\mathcal{L}_1\rho.\label{eq:multibandmain}
\eea
The term describing scattering of laser photons, denoted $\mathcal{L}_1\rho$, can be written in this basis as
\bea
\mathcal{L}_1\rho&=&-\frac{1}{2}\sum
\gamma_{\vh{i},\vh{j}}^{(\vh{k},\vh{l},\vh{m},\vh{n})}\comm{b_{\vh{i}}^{(\vh{k})\,\dag}b_{\vh{i}}^{(\vh{l})}}
{\comm{b_{\vh{j}}^{(\vh{m})\,\dag}b_{\vh{j}}^{(\vh{n})}}{\rho}}. \label{eq:multibanddissipative}
 \eea
Here the sum runs over $\{\vh{i},\vh{j}, \vh{k},\vh{l}, \vh{m},\vh{n}\}$ and the matrix elements for the different processes are defined as:
\bea
\gamma_{\vh{i},\vh{j}}^{(\vh{k},\vh{l},\vh{m},\vh{n})}&=&\iint d^3xd^3y\frac{\Gamma\Omega(\vh{x})\Omega^{\ast}(\vh{y})}{4\Delta^2}F(k_{eg}(\vh{x}-\vh{y}))
w_{\vh{i}}^{(\vh{k})}(\vh{x})w_{\vh{i}}^{(\vh{l})}(\vh{x})w_{\vh{j}}^{(\vh{m})}(\vh{y})w_{\vh{j}}^{(\vh{n})}(\vh{y}).\label{eq:multibandlast}
\eea
\end{widetext}

\emph{Single-band Hubbard model} -- For red-detuning, where the dominant processes in the scattering will return atoms to the lowest band, it can be convenient to focus on the physics resulting from these dominant processes. If, in addition, we neglect the small terms that are not diagonal in a position basis ($\sim F(k_{e,g}a)$), then \eqref{eq:master_equation} reduces to
\be\label{eq:QT_master_eq}
\dot{\rho}=-i[H_{BH},\rho]+\sum_{i}\gamma (n_i\rho n_i-1/2n_i n_i\rho-1/2\rho n_i n_i).
\ee
Here $\gamma$ is the effective scattering rate, and $H_{BH}$ is the Bose-Hubbard Hamiltonian for the lowest band,
as given in Eq.~(\ref{eq:HBH})
with $b_i\equiv b_i^{(\mathbf{0})}$, $n_i=b_i^{\dag}b_i$, and $\varepsilon\equiv \varepsilon^{(\vh{0})}=0$. This form of the master equation is used below for quantum trajectories simulations, and also clearly demonstrates the localizing effects of spontaneous emissions that are discussed in the following subsection. The expressions here for $U$ and $J$ are also used when describing the initial state of the system, where atoms are assumed to be confined to the lowest Bloch band.

\subsection{Localization of atoms due to spontaneous emission events}
\label{sec:localisation}

\begin{figure}[tb]\centering
\includegraphics[width=8cm]{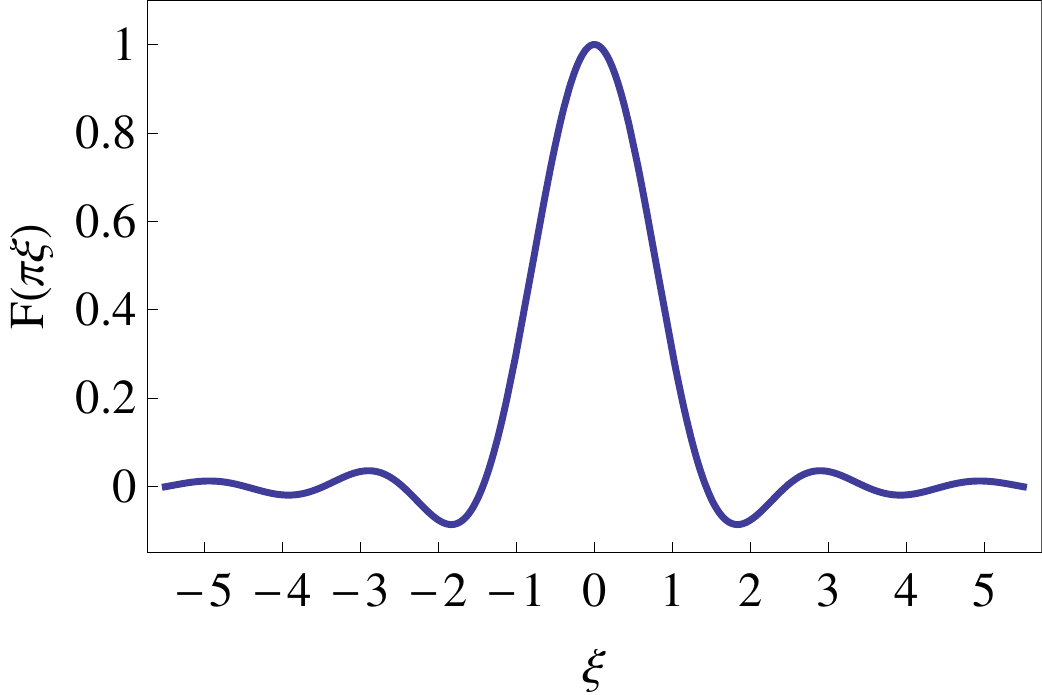}
\caption{Plot of the function $F(\vh{\xi})$ for $\vh{\xi}$ parallel to $\vh{d}$ (see Eq.~(\ref{eq:F_def}) for the definition of $F$). Note that the localized form of this function reflections the localization of atoms in spontaneous emission events.}\label{fig:F_plot}
\end{figure}

As mentioned above, a key characteristic of spontaneous emission events, both for single particle and many-particle systems, is that they localize the atom undergoing the scattering event on a length scale given by the wavelength of the emitted photon \cite{Hol96}. This is most clearly seen from the recycling term in Eq.~(\ref{eq:dissipation}), where the function $F(k_{eg}(\vh{x}-\vh{y}))$ determines the coherence between emissions at different points in space, $\vh{x}$ and $\vh{y}$. In Fig.~\ref{fig:F_plot} we plot the function $F(\vh{\xi})$, showing that it is clearly localized. The length scale of the localization is then determined by $1/k_{eg}$, which is proportional to the wavelength of emitted light. When we write the master equation in a Wannier basis, the fact that $F(k_{eg}(\vh{x}-\vh{y}))$ is localized on a length scale similar to the lattice period means that the master equation will be approximately diagonal in position space, with the recycling term given by
\begin{align*}
\mathcal{J}\rho\approx&\sum_{\vh{k},\vh{l},\vh{m},\vh{n},\vh{i}}\gamma_{\vh{i},\vh{i}}^{(\vh{k},\vh{l},\vh{m},\vh{n})}b_{\vh{i}}^{(\vh{k})\,\dag}b_{\vh{i}}^{(\vh{l})}\rho b_{\vh{i}}^{(\vh{m})\,\dag}b_{\vh{i}}^{(\vh{n})}.
\end{align*}
This diagonal form in $b_i^{\dag} b_i$ will dephase coherent superpositions in which atoms are delocalized over many sites into mixtures in which for each classical possibility the atom is localized. This is a representation of the fact that information about the atom's position is transmitted to the environment by the emitted photon, and in the absence of a measurement of the photon, we are left with a classical mixture of different possible locations where the photon could have been scattered. Below we will see that this mechanism is very important when one considers the effect of spontaneous emissions on many-body states. In particular, this localization tends to destroy long-range order, which is a key property of superfluid states.

\section{Quantifying Decoherence and Heating}\label{sec:quantification}

We start our discussion of decoherence and heating in subsection \ref{sec:perturbation_theory} by studying the evolution of different many body states in perturbation theory as described by the master equation \eqref{eq:master_equation}. The calculations are based on the full multi-band master equation, but use perturbation theory in the limit of small $\gamma/J$, $\gamma/U$, where $\gamma$ is the effective scattering rate, and also neglect interactions of atoms that have undergone spontaneous emission events. We quantify heating for different many-body states, and for different signs of the laser detuning. We first investigate the effective scattering rate, and the rate of energy increase, and find results that are very similar to the single-particle case presented in Sec.~\ref{sec:singleparticle}, but which are modified quantitatively due to collective effects. However, because atoms scattered to higher bands will typically not thermalize with those in the lowest band on experimental timescales, the increase in mean energy is not sufficient to determine the change in the many-body state due to heating. Instead, it is important to directly investigate changes in the characteristic correlation functions. We investigate the change in correlation functions in subsections \ref{sec:quantspdm} and \ref{propagation}.

In Sec.~\ref{propagation} we combine quantum trajectories techniques with t-DMRG methods in order to account for longer time evolutions and thermalization after spontaneous emission events by propagating the master equation directly in time. However, for these purposes we find it convenient to restrict calculations to the lowest band master equation \eqref{eq:QT_master_eq}, relevant for deep red detuned lattices. This gives us an exact solution, but only for one dimensional lattices.

Finally in subsection \ref{sec:gutzwiller} we present a mean field description based on Gutzwiller mean field theory to investigate the master equation \eqref{eq:master_equation} for a 3D lattice. We expect this approach to provide a semi-quantitative description for the heating processes, but this will certainly not capture details for the spatial dependence of the first order correlation functions, and their time dependence due to heating.

\subsection{Perturbation Theory}\label{sec:perturbation_theory}
\subsubsection{Scattering Rate}

\begin{figure}[tb]\centering
\includegraphics[width=8cm]{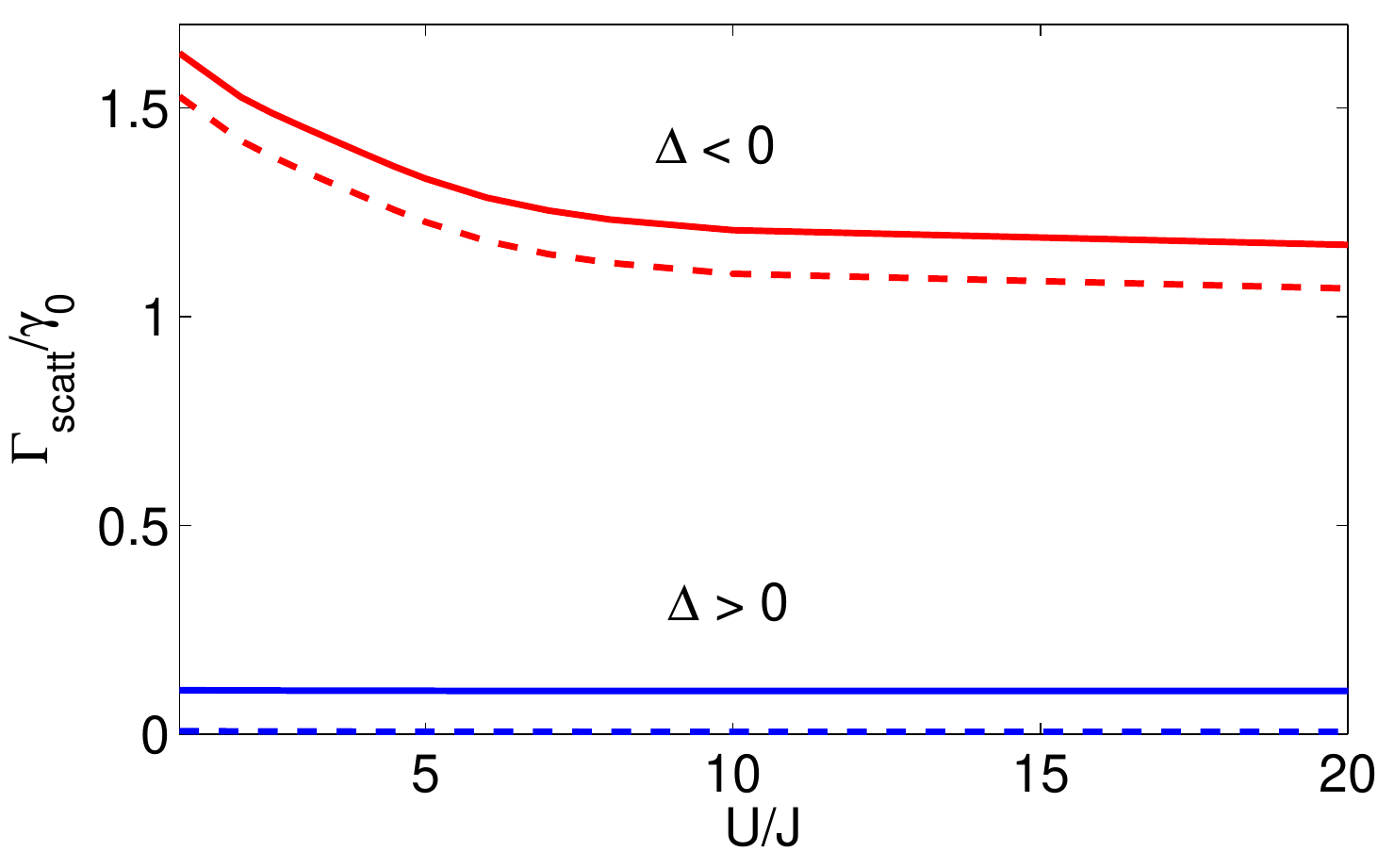}
\caption{(color online) Comparison between the total process rate $\Gamma_{\rm scatt}
/\gamma_0$ (solid lines) and the rate for processes back to the lowest band (dashed lines) for $\Delta<0$ (upper lines) and a $\Delta>0$ (lower lines) in the ground state of a 1D lattice (Using DMRG ground states for Bosons in a 1D lattice: $V_x=V_y=30 \,E_R, V_z=10 \,E_R$) The rates include scattering from all three lattice generating beams.}\label{fig:scattering_rate}
\end{figure}

We can determine the scattering rate for spontaneously emitted photons directly from Eq.~\eqref{eq:master_equation}, and we find
\begin{widetext}
\begin{align}
\Gamma_{\rm scatt}=&\frac{\Gamma}{4\Delta^2}\iint d^3x d^3yF(k_{eg}(\vh{x}-\vh{y}))\Omega(\vh{x})\Omega(\vh{y})
\mathrm{Tr}\left\{\hat{\psi}^{\dag}(\vh{x})\hat{\psi}(\vh{x})\rho\hat{\psi}^{\dag}(\vh{y})\hat{\psi}(\vh{y})\right\}\no\\
=&\frac{\Gamma}{4\Delta^2} \int d^3x\Omega(\vh{x})^2\mathrm{Tr}\left\{\hat{\psi}^{\dag}(\vh{x})\hat{\psi}(\vh{x})\rho\right\}
+\frac{\Gamma}{4\Delta^2} \iint d^3x d^3yF(k_{eg}(\vh{x}-\vh{y}))\Omega(\vh{x})\Omega(\vh{y})
\mathrm{Tr}\left\{\hat{\psi}^{\dag}(\vh{y})\hat{\psi}^{\dag}(\vh{x})\hat{\psi}(\vh{y})\hat{\psi}(\vh{x})\rho\right\}. \label{eq:scattrate}
\end{align}
\end{widetext}
As in the single particle case discussed in Sec.~\ref{sec:heatredblue}, this rate depends strongly on the sign of the detuning, and it now also depends strongly on the characteristic many-body state through the correlation functions computed for $\rho$. In Fig.~\ref{fig:scattering_rate} we plot the scattering rate for atoms confined to move in 1D by an anisotropic lattice, for which we can compute the many-body ground state of the Hubbard model using t-DMRG methods, and hence the correlation functions that appear in Eq.~\eqref{eq:scattrate}. The scattering rate is plotted as a function of $U/J$ for a system at unit filling, and we note immediately that the scattering rate is again much larger for red-detuned light $\Delta<0$ than for blue-detuned light $\Delta>0$. The scattering the red-detuned case is again dominated by processes returning atoms to the lowest Bloch band, as shown by the dashed lines in Fig.~\ref{fig:scattering_rate}. In the superfluid regime (i.e. for $U/J\lesssim 3.37$), the rate for processes that leave the atom in the lowest band is increased by bosonic enhancement \cite{Goe01}, due to the presence of other delocalized atoms at the place where the spontaneous emission event occurs. Since the process that returns an atom to the lowest band is much more prominent for red-detuned light, this enhancement is much more visible in that case than for blue-detuned light. As the interaction strength is increased, atoms tend to be exponentially localized on individual sites in the Mott-Insulator phase, corresponding to $U/J\gtrsim 3.37$, and for increasing $U/J$ this enhancement is strongly reduced.

\subsubsection{Total rate of energy increase}\label{sec:total_heating}
From the master equation Eq.~(\ref{eq:master_equation}), we can calculate
the rate of change $d\langle  H \rangle/dt$ of the total mean
energy of the atom cloud due to scattering of laser photons. Here $H$ is the hermitian part of the effective Hamiltonian in equation (\ref{eq:master_equation}), and $\mathcal{L}_1\rho$ is the part of the master equations that describes incoherent scattering of the laser photons:
\begin{align}\label{eq:energy_increase}
\frac{d}{dt}\langle H \rangle=\mathrm{Tr}\{ H \mathcal{L}_1\rho\}.
\end{align}
For each standing waves with laser wavenumber $k_l$, giving an effective Rabi frequency $\Omega(x)=\Omega_0\cos(k_l x)$ we find (see Appendix \ref{ap:heting_rate})
\begin{align}\label{eq:energy_increase_result}
\frac{d}{dt}\langle  H \rangle=\frac{\Gamma |\Omega_{0}|^2}{4\Delta^2} E_R N.
\end{align}
Thus, the total rate of energy increase is \emph{independent of the sign of the detuning, and of the properties of the many-body state}. Furthermore, the increase in \emph{energy per particle} is the same as in the \emph{single particle case} presented in Sec.~\ref{sec:heatredblue}.
However, as we will discuss in the next subsection, this is not the key quantity for determining changes to the many-body state.

\subsubsection{Thermalization and atoms in excited Bloch bands}\label{sec:lack_of_thermalization}
The dominant contribution to the energy computed in Sec.~\ref{sec:total_heating} is due to atoms scattered into higher Bloch bands, which can easily be shown in the Lamb Dicke limit (see Sec.~\ref{sec:heatredblue}). The energy gain in a scattering event in which the atom remains in the lowest band is on the order of the tunneling rate $J$, whereas atoms scattered into higher bands gain an energy greater than the bandgap energy $\varepsilon_{\rm gap}$. However, because $\varepsilon_{\rm gap}\gg J,U$, it is not possible for these atoms that have been scattered to higher bands to thermalize this energy input on typical experimental timescales. Indeed. it would require very high-order processes in perturbation theory in $J/\varepsilon_{\rm gap}$ and $U/\varepsilon_{\rm gap}$ in order to return these atoms to the lowest band. As a result, these processes that create a large change in the mean energy of the state will nonetheless often contribute very little to a change in the many-body state. Thus it is clear that the total energy \emph{does not} completely characterize the change in the many-body state. Instead, we must look directly at other quantities, such as correlation functions determining the characteristic properties of the many-body state, and how they are changed as a result of the heating processes. We will show below that although the rate of energy increase is independent of the sign of the detuning, that the change in character of the many-body state can be strongly dependent on that sign.

\subsubsection{Single particle density matrix and momentum distribution}

\label{sec:quantspdm}

\begin{figure}[tb]
  \centering
\includegraphics[width=8cm]{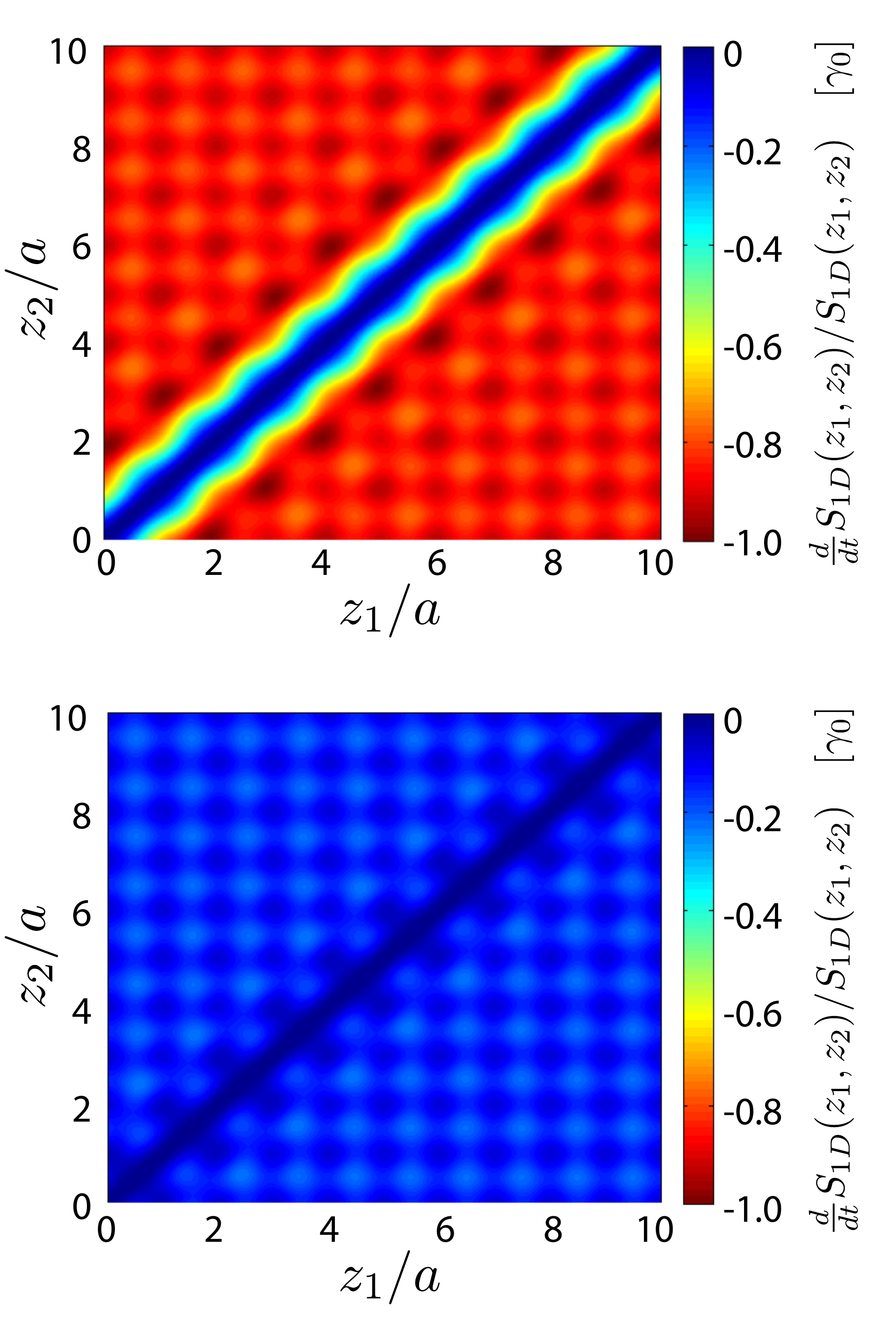}
 \caption{(Color online) Relative rate of change of the (integrated) single particle density matrix $S_{1D}(z_1,z_2)=\iint dxdy\langle\hat{\psi}^{\dag}(x,y,z_1)\hat{\psi}(x,y,z_2)\rangle$ in an effectively 1D lattice ($V_x=V_y=30\,E_R$; $V_z=10\,E_R$). In the tightly bound transversal directions the atoms are assumed to be in the lowest band. Scattering from all three lattice generating laser beams is taken into account (with weights corresponding to the lattice depths). (The lattice constant is denoted by $a$.)
}
 \label{fig:spdm_decay_rel}
\end{figure}

For bosons in an optical lattice, a key quantity in characterizing the many body state is the single particle density matrix $S(\vh{x_1},\vh{x_2})=\langle\hat{\psi}^{\dag}(\vh{x_1})\hat{\psi}(\vh{x_2})\rangle$. In the superfluid regime, the system exhibits off-diagonal long range order (or quasi-long range order for 1D systems), whereas in the Mott Insulator regime this function decays exponentially as a function of distance $|\vh{x_1}-\vh{x_2}|$. In first order perturbation theory in $\gamma_0/J$, $\gamma_0/U$, the change of the single particle density matrix due to scattering of laser photons is given by
\begin{align}\label{eq:spdm_change}
\dot S(\vh{x_1},\vh{x_2})=&\mathrm{Tr}\{\hat{\psi}^{\dag}(\vh{x}_1)\hat{\psi}(\vh{x}_2)\mathcal{L}_1\rho\}\\
=&-\frac{1}{2}\frac{\Gamma}{4\Delta^2}
\left(\Omega(\vh{x}_1)^2+\Omega(\vh{x}_2)^2\right.\nonumber\\
&\left.-2F(k(\vh{x}_1-\vh{x}_2))\Omega(\vh{x}_1)\Omega
(\vh{x}_2)\right)\langle\hat{\psi}^{\dag}(\vh{x}_1)\hat{\psi}(\vh{x}_2)\rangle.
\end{align}
Here we have also neglected interactions with atoms that have undergone spontaneous emissions. These will be treated in Sec.~\ref{propagation}. Noting that $F(k(\vh{x}_1-\vh{x}_2))\rightarrow 0$ for $ |\vh{x}_1-\vh{x}_2|\gg a$ the off-diagonal long range order changes according to:
\begin{align}\label{eq:spdm_offdiag_cahnge}
&\mathrm{Tr}\{\hat{\psi}^{\dag}(\vh{x}_1)\hat{\psi}(\vh{x}_2)\mathcal{L}_1\rho\}
\xrightarrow{|\vh{x}_1-\vh{x}_2|\gg a}\nonumber\\
&-\frac{1}{2}\frac{\Gamma}{4\Delta^2}
\left(\Omega(\vh{x}_1)^2+\Omega(\vh{x}_2)^2\right)
\langle\hat{\psi}^{\dag}(\vh{x}_1)\hat{\psi}(\vh{x}_2)\rangle.
\end{align}

The relative rates of change for the single particle density matrix as a function of $\vh{x_1}$ and $\vh{x_2}$ are plotted in Fig.~\ref{fig:spdm_decay_rel} for red and blue detuned light, normalized to $\gamma_0$. We see a clear difference between the results for red and blue detuning. The reason for this is that the breakdown of long-range correlations is rooted in the localization effect of spontaneous emission events, as was discussed in Sec.~\ref{sec:localisation}. This localization depends not on the rate of energy input into the system, but rather on the total scattering rate, which is larger in the red-detuned case. Hence, the breakdown occurs substantially faster for the red-detuned lattice than for blue detuning.

At the same time, the localization effect is much more harmful for a superfluid state in the weakly interacting limit, than for a Mott Insulator state, where the particles are already exponentially localized at different sites. In the extreme limit $U/J \rightarrow \infty$, the only significant change in the state is the (relatively rare) transfer of some atoms to higher bands.

\begin{figure}[tb]
  \centering
  \includegraphics[width=8cm]{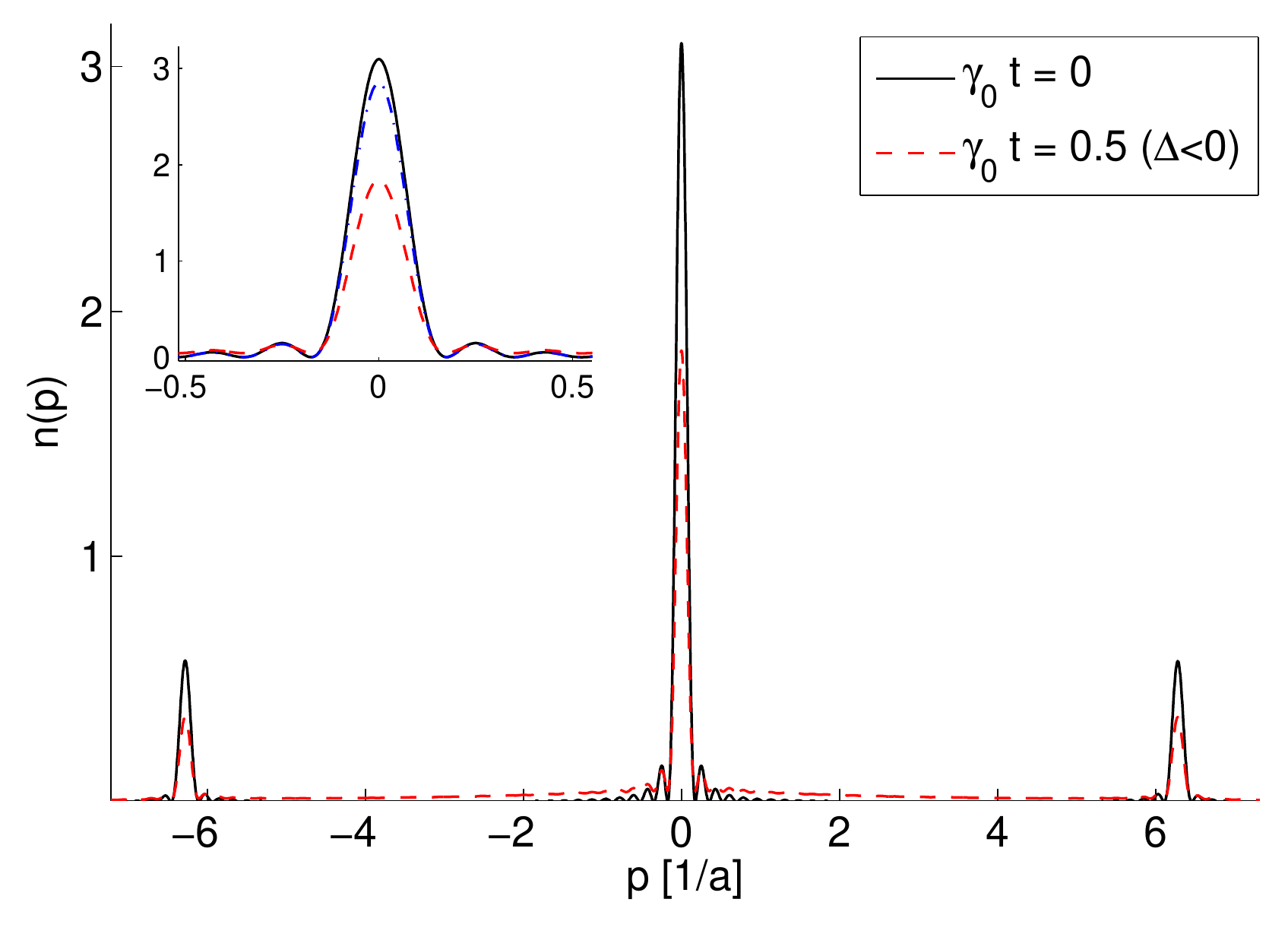}
\caption{(Color online) Momentum distribution (in arbitrary units) of the ground state of noninteracting particles in a 1D lattice of depth $V = 10 E_R$ (black), and after a time $\gamma_0t=0.5$ if the lattice is generated by a red laser (red, dashed). The inset shows the central peak initially (black, solid) and after $\gamma_0t=0.5$ in a red (red, dashed) and in a blue lattice (blue, dashed-doted).   (36 particles on 36 sites)}\label{fig:momentum_perturbation_theory}
\end{figure}

Note that the single particle density matrix is also directly connected to the momentum distribution $n(\vh{p})$, meaning that the measurements that are made in experiments (including the comparison made with quantum monte-carlo calculations in Ref.~\cite{Tro09}) are directly related to the changes in these correlation functions. Specifically,
\be
n(\vh{p})=\frac{1}{(2\pi)^3}\iint d^3x_1d^3x_2 S(\vh{x}_1,\vh{x}_2)e^{-i\vh{p}\cdot(\vh{x}_1-\vh{x}_2)},
\ee
so that the presence of (quasi-) off-diagonal long range order in the superfluid regime gives rise to peaks at reciprocal lattice vectors. The breakdown of long-range order due to spontaneous emission events is then directly related to a decrease in the visibility of these peaks, as we show in Fig.~\ref{fig:momentum_perturbation_theory}). Again, in the extreme Mott Insulator limit, the state is very insensitive to spontaneous emissions, with the initial momentum distribution being given by the Fourier transform of the lowest band Wannier function, and changing only due to processes where an atom is scattered into higher bands. Since the timescales for these processes are similar for red and blue detuning, both lattices have similar effect on the Mott Insulator.

Using perturbation theory calculations it is difficult to quantify the effects of interactions between particles on the lattice after the spontaneous emission has occurred, in particular effects due to partial thermalization of the energy added to the system. In order to address this, it is necessary to find a method to propagate the master equation directly in time. This is discussed in the next section.

\subsection{Propagation of the single band master equation (\ref{eq:QT_master_eq}) with Quantum Trajectories \& TEBD}
\label{propagation}
In order to further quantify the heating process, especially the effects of interactions in the system after spontaneous emission events have occurred, we now investigate means to directly integrate the master equation (\ref{eq:master_equation}). For 1D systems we can compute  time-dependent expectation values from the master equation exactly by combining time-dependent density matrix renormalization group (t-DMRG) methods with quantum trajectory techniques, as discussed in Ref.~\cite{Dal09}. In order to simplify the numerical computation, we focus on processes returning atoms to the lowest Bloch band, which are dominant for red detuning, as discussed above, and base the simulations on the single-band master equation given in Eq.~\eqref{eq:QT_master_eq}. In the 2D/3D case, it is numerically prohibitively expensive to apply these methods directly for realistic system sizes, however we expect that certain aspects can be described semi-quantitatively by a mean-field theory treatment similar to that in Ref.~\cite{Die10}, which will be discussed in Sec.~\ref{sec:gutzwiller}.
\label{sec:QT_TEBD}

In the combination of quantum trajectories and t-DMRG, t-DMRG \cite{Vid03,Dal04,Whi04} provides a convenient means to propagate states that are not too far from equilibrium in a 1D system, whereas quantum trajectories \cite{trajectoriesc,trajectoriesg} is a method to compute time-dependent correlation functions from the master equation based on a stochastic propagation of states. Briefly, the idea is that each stochastic trajectory begins from an initial pure state (sampled from the initial density matrix), and is propagated based on the non-Hermitian Hamiltonian $H^{QT}_{\rm eff}=H_{BH}-(\gamma/2)\sum_i n_i n_i$, except for at randomly sampled times $t_j$, where \textit{quantum jumps} occur,
\begin{equation}
{i} \frac{d}{dt}\ket{\psi(t)}=H_{\rm eff} \ket{\psi(t)}; \,\,\,\ket{\psi(t_j^+ )}=\frac{n_{i_j}\ket{\psi(t_j)}}{||n_{i_j}\ket{\psi(t_j)}||}, \label{trajevolution}
\end{equation}
corresponding to the localization of a particle onsite due to the spontaneous emission event. In the stochastic simulation the times $t_j$ are points where the norm of the state falls below a randomly chosen threshold. At these times, a random site $i_j$ is selected for the spontaneous emission event according to the probabilities $p_{i_j}\propto \bra{\psi(t_j)} n_{i_j}^\dag  n_{i_j}\ket{\psi(t_j)}$ and applied. Expectation values are computed based on stochastic averages, which converge rapidly as a function of the number of trajectories. Each of the computations presented here was based on sampling 1000 trajectories, which was possible on a timescale of a few days with a cluster of ca.~100 CPU cores. In every case, numerical convergence was checked, with the number of states retained for decompositions in t-DMRG being computed to $\chi=200$.

\begin{figure}[tb]
  \centering
  \includegraphics[width=8cm]{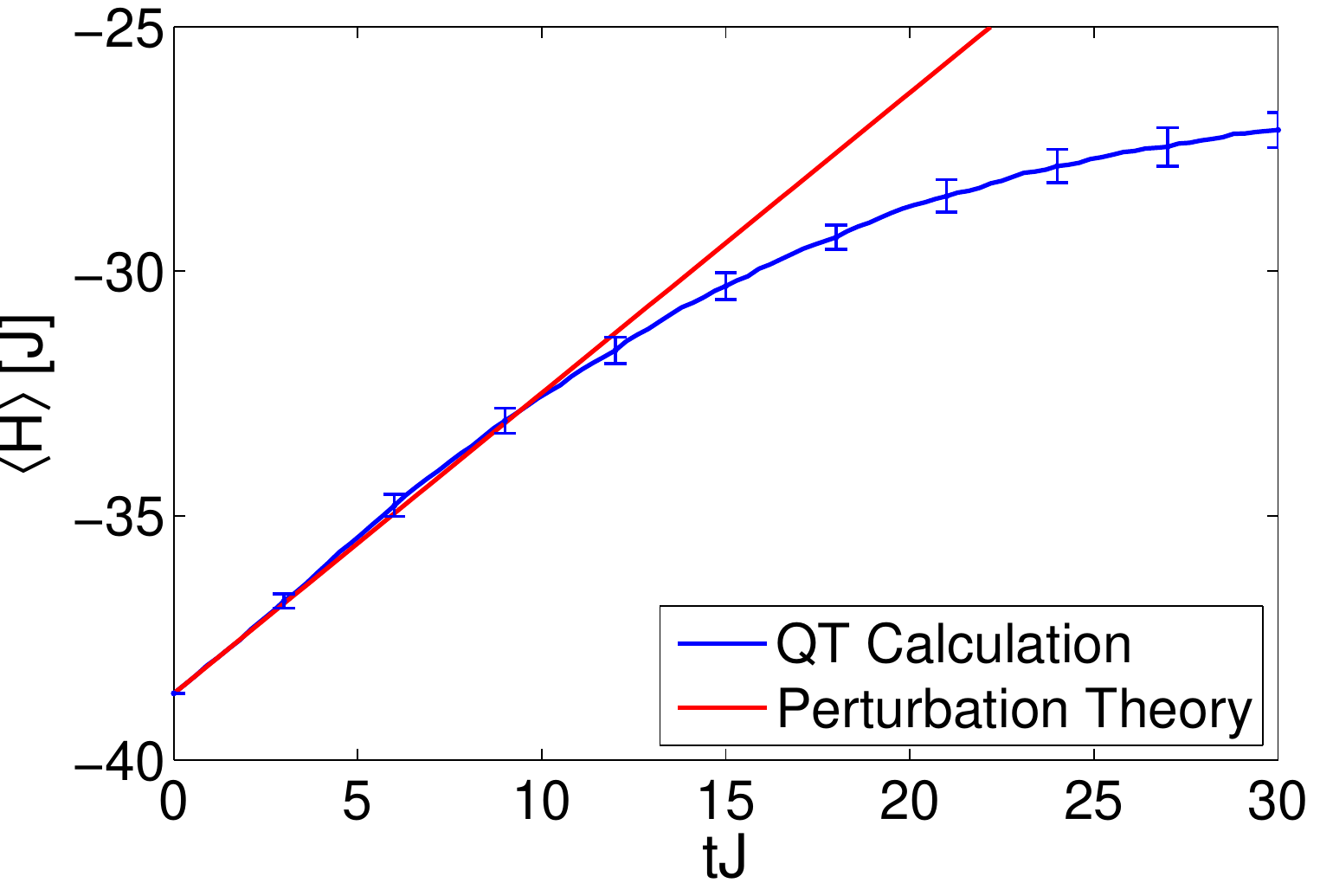}
  \caption{Development of the mean energy in a system with 36 particles on 36 sites, at $U/J=3$ just below the superfluid to Mott Insulator transition and $\gamma=0.01J$, computed from Eq.~\eqref{eq:QT_master_eq} via quantum trajectories methods, compared with the equivalent result from first order perturbation theory. Statistical error bars are shown based on the stochastic average over 1000 trajectories.} \label{fig:Energy_TEBD}
\end{figure}

\begin{figure}[tb]
  \centering
  \includegraphics[width=8cm]{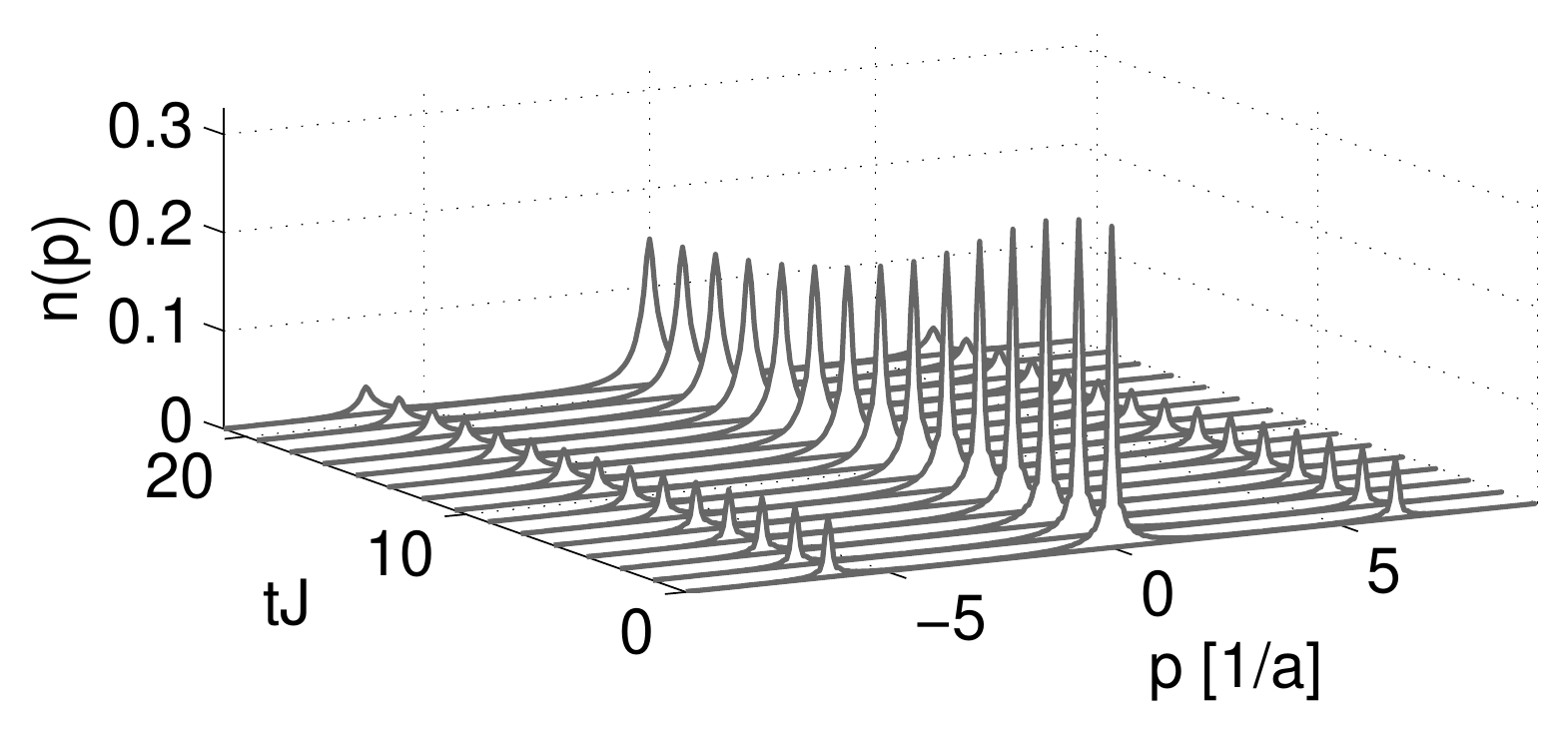}
  \caption{Momentum distribution (in arbitrary units) calculated for a 1D lattice (36 particles on 36 sites) with $U/J=3$ (i.e. in the superfluid regime) and a scattering rate $\gamma=0.01J$ from Eq.~\eqref{eq:QT_master_eq} via quantum trajectories methods in a lattice of depth $V = 10 E_R$. Averages are taken over 1000 trajectories. The scale of the momenta axis is $1/a$, where $a$ is the lattice constant.}\label{fig:momentum_time}
\end{figure}

\begin{figure}[tb]
  \centering
  \includegraphics[width=7cm]{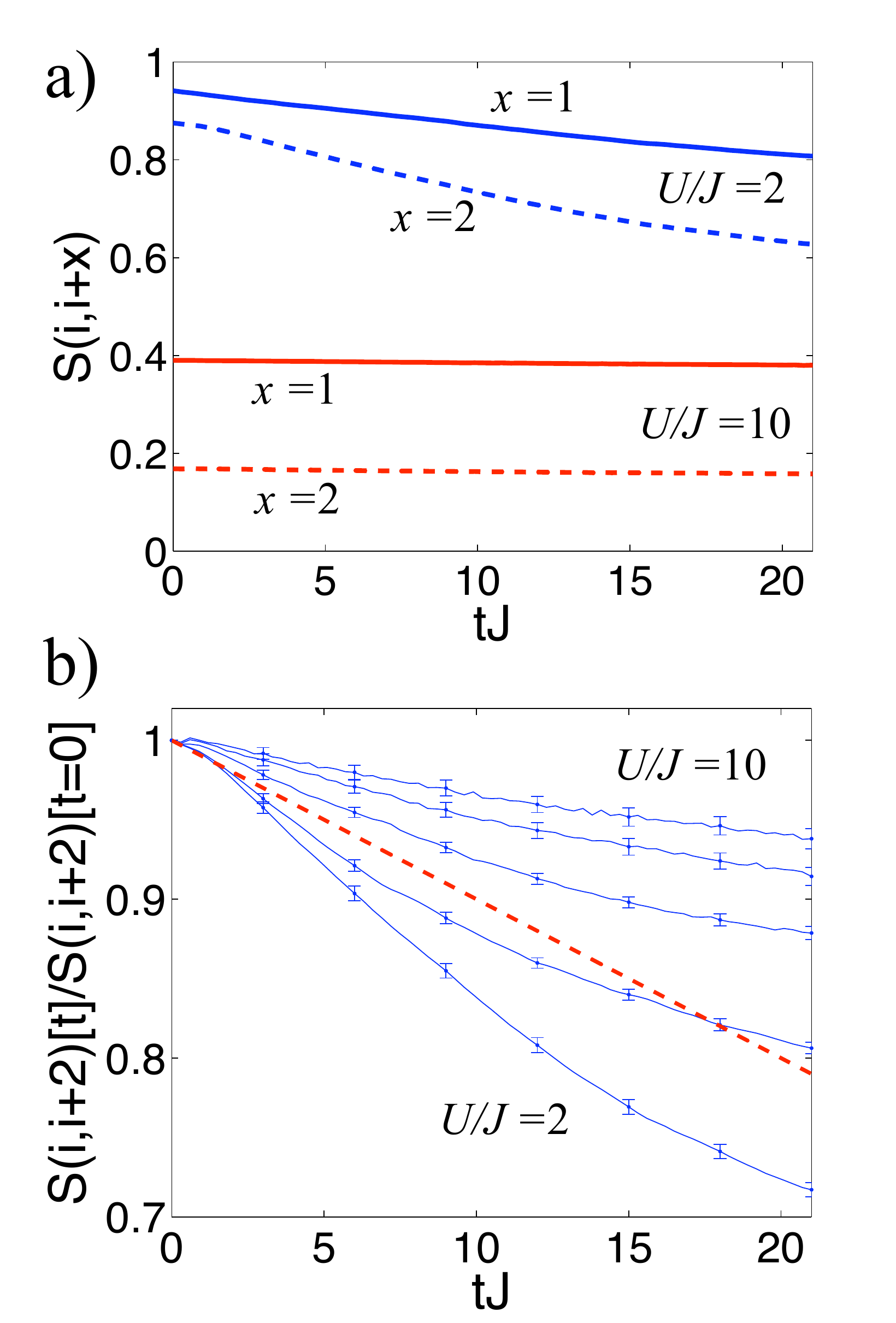}
  \caption{Comparison of the decay of off-diagonal correlations in the single particle density matrix $S(i,j)=\langle b_i^\dag b_j  \rangle$ as a function of time. These are computed from the master equation for the lowest band Eq.~\eqref{eq:QT_master_eq} with $\gamma/J=0.01$ for 24 particles on a 24 site lattice by combining quantum trajectories methods with t-DMRG. (a) Comparison of the decay of off-diagonal elements $\sum_i S(i,i+x)$ when $U/J=2$ (superfluid regime, upper lines) and $U/J=10$ (Mott Insulator regime, lower lines). In each case we show correlations at separation distances $x=1$ (solid lines) and $x=2$ (dashed lines), each of which are averaged over $i$. We see that the correlations for the Mott Insulator state at $U/J=10$ are almost constant, whereas for $U/J=2$, the decay becomes more rapid than at initial times.  (b) Comparison of $S(i,i+2)$ averaged over $i$ and normalized to the value at time $t=0$,  $|S(i,i+2)|[t=0]$ for $U/J=2,4,6,8,10$ (seen here from bottom to top). The dashed line shows the corresponding result from perturbation theory. Each computation was averaged over 1000 trajectories, and error bars are shown in (b). For (a), the statistical errors fit inside the line thickness.} \label{fig:spdm_decay_TEBD}
\end{figure}

Results from these calculations are shown in Figs.~\ref{fig:Energy_TEBD}, \ref{fig:momentum_time} and \ref{fig:spdm_decay_TEBD}. In Fig.~\ref{fig:Energy_TEBD}
we show the increase in energy for a system with 36 particles on 36 sites interaction strength $U/J=3$ and $\gamma=0.01J$. We observe excellent agreement with results from perturbation theory at short times, where we have the linear relation $\langle H_{BH} \rangle_t\approx \langle H_{BH} \rangle_0+\frac{d}{dt}\langle H_{BH} \rangle|_{t=0}t$, with the rate of change of the mean energy from Eq.~(\ref{eq:QT_master_eq}) given by $\frac{d}{dt}\langle H_{BH} \rangle|_{t=0}=\gamma J\sum_{\langle i,j\rangle}\langle b_i^{\dag}b_j\rangle$. In Fig.~\ref{fig:momentum_time}, we similarly see the characteristic decay of the central peak in the momentum distribution as off-diagonal correlations decay over time.

In Fig.~\ref{fig:spdm_decay_TEBD}, we show the evolution of the off-diagonal elements of the correlation functions $S(i,j)=\langle b_i^{\dag}b_{j}\rangle$, where in contrast to the results presented in the previous section, we now include the effects of collisions after spontaneous emission events, i.e., partial thermalization of the energy added to the system. The basic change to these previous results is immediately apparent in Fig.~\ref{fig:spdm_decay_TEBD}a, which shows off-diagonal correlations in the superfluid regime for $U/J=2$ and in the Mott Insulator regime for $U/J=10$ as a function of time. Where the perturbation theory results, neglecting scattering after a decay event, give us identical rates of relative decay of these values for different interaction strenghts, here the rates depend strongly on $U/J$. To further quantify this, we plot values of the correlation function normalized to the same initial value in Fig.~\ref{fig:spdm_decay_TEBD}b, for different values of $U/J$ ranging between $U/J=2$ and $U/J=10$. From perturbation theory calculations not including collisions between atoms after spontaneous emission events, these should all decay at the same rate, as $\frac{d}{dt}\langle b^\dag_{\vh{i}} b_{\vh{j}}\rangle=-\gamma \langle b^\dag_{\vh{i}} b_{\vh{j}}\rangle$. However, we see that in the superfluid case, the rates decay more rapidly at intermediate times, as thermalization creates further decay of the off-diagonal correlations in addition to the initial localization effect of the spontaneous emission events. In contrast, for the Mott-Insulator limit, the results deviate already strongly from the perturbation theory results on a timescale given by $1/U$, and on longer timescales the off-diagonal elements are barely changed, as the strong interactions tend to result in the small local correlations being reestablished. It is worthwhile to note that longer-range correlations in the Mott-Insulator state can actually slightly increase in comparison with their initially exponentially small values, as energy added to the system is thermalized.
\subsection{Propagation of the master equation via a mean-field ansatz}
\label{sec:Gutzi}
\label{sec:gutzwiller}
In two or three dimensions, exact computation of time-dependent dynamics becomes numerically extremely expensive for all but very small system sizes. However, it is possible to gain insight from the time-evolution based on a mean-field ansatz. Below we employ a treatment of the master equation analogous to the Gutzwiller mean-field analysis of the ground states of interacting systems (valid in higher dimensions), where we generalize these ideas to time-dependent dynamics. Such a Gutzwiller mean field approach has been developed in Ref.~\cite{Die10} to describe dynamical quantum phase transitions as a competition between coherent Hamiltonian and incoherent Liouvillian dynamics. This method is simple to apply, and allows us to easily include higher bands in the computation, which can be numerically expensive for the exact methods discussed in the previous section. A simple picture of the heating can be built up in terms of the distribution of particles in different bands and the increase in energy and entropy in the system, as well as decay of the superfluid order parameter due to heating, if we begin in a superfluid state. The inclusion of higher bands here also allows us to quantitatively see the lack of thermalization of energy input as atoms transferred to higher bands. As with ground state calculations based on a typical Gutzwiller ansatz, the results are expected to be more accurate for systems in higher dimensions, and it is not possible to extract accurate information about the spatial dependence of correlation functions, which we have already discussed in some detail for the 1D case.

For zero temperature, ground state calculations of the single band Bose Hubbard Hamiltonian Eq.~(\ref{eq:HBH}), the Guzwiller ansatz assumes a product state over the lattice sites,
\begin{equation}
\ket \Psi=\Pi_{i}\ket{\phi_i}=\prod_i\sum_{n}f_n^{(i)}\ket{n}_i.
\end{equation}
The amplitudes $f_n^{(i)}$ of states with occupation number $n$ at site $i$ are used as variational parameters to minimize $\langle H_{BH}\rangle$ with the constraints of normalized wave function and a fixed mean number of particles per lattice site.
Here we identify the mean field Hamiltonian, $$H_{MF}=\sum_i\lr{-J\lr{\psi_ib_i^{\dag}+\psi_i^{\ast}b_i}+\frac{1}{2} Ub_i^{\dag\,^2}b_i^2},$$ as a sum of local operators. The superfluid phase manifests itself in a nontrivial superposition $\ket{\phi_i}=\sum_{n}f_n^{(i)}\ket{n}_i$, with a nonzero expectation value of the destruction operator $\psi_i\equiv \sum_{j|i}\langle\phi_j|b_j|\phi_j\rangle$, whereas in the Mott Insulator phase with filling $n_0$ one has $f_n^{i}=\delta_{n,n_0}$, such that $\psi_i=0$. At unit filling this transition from superfluid to Mott phase occurs at $U/(zJ)\approx5.8$, where $z$ denotes the number of nearest neighbors.

In the same spirit we use a factorization ansatz to solve our master equation \eqref{eq:multibandham}--\eqref{eq:multibandlast}:  we write $\rho=\bigotimes_i\rho_i$ with $\rho_i=\rm{Tr}_{\neq l}\{\rho\}$ \cite{Die10}, and derive an equation of motion for the density operator at site $i$: $\dot\rho_i=\rm{Tr}_{\neq i} \{\mathcal{L}\rho\}$.
Then the multiband equation of motion is given by
\begin{align}\label{eq:mean_field_master_eq}
\dot{\rho}_{\vh{i}}=-i\comm{h_{MF,\vh{i}}}{\rho_{\vh{i}}}+\mathcal{L}_{MF,\vh{i}}\rho_{\vh{i}}.
\end{align}
The multiband mean-field Hamiltonian at lattice site $\vh{i}$ is:
\begin{widetext}
\begin{align}
h_{MF,\vh{i}}&=& \sum_{\vh{j}|\vh{i},\vh{n}}\lr{-J_{\vh{i},\vh{j}}^{(\vh{n})}\lr{\mean{b_{\vh{j}}^{(\vh{n})\,\dag}} b_{\vh{i}}^{(\vh{n})}+\mean{ b_{\vh{j}}^{(\vh{n})}}b_{\vh{i}}^{(\vh{n})\,\dag}}+\varepsilon^{(\vh{n})}b_{\vh{i}}^{(\vh{n})\,\dag} b_{\vh{i}}^{(\vh{n})}}+\sum_{\vh{n},\vh{m},\vh{k},\vh{l}}\frac{1}{2}U^{(\vh{k,l,m,n})}b_{\vh{i}}^{(\vh{k})\,\dag}
b_{\vh{i}}^{(\vh{l})\,\dag}b_{\vh{i}}^{(\vh{m})}b_{\vh{i}}^{(\vh{n})},
\end{align}
where the brackets denote the expectation value under the density operator $\rho$: $\mean{\dots}\equiv\rm{Tr}\{\dots\rho\}$. The incoherent part at lattice sites $\vh{i}$ is:
\bea
\mathcal{L}_{MF,\vh{i}}\rho_{\vh{i}}=\sum_{\vh{n},\vh{m},\vh{k},\vh{l}}\gamma_{\vh{i},\vh{i}}^{(\vh{k},\vh{l},\vh{m},\vh{n})}
\left(b_{\vh{i}}^{(\vh{k})\,\dag}b_{\vh{i}}^{(\vh{l})}\rho_{\vh{i}} b_{\vh{i}}^{(\vh{m})\,\dag}b_{\vh{i}}^{(\vh{n})}-
\frac{1}{2} b_{\vh{i}}^{(\vh{m})\,\dag}b_{\vh{i}}^{(\vh{n})}b_{\vh{i}}^{(\vh{k})\,\dag}b_{\vh{i}}^{(\vh{l})}\rho_{\vh{i}}-
\frac{1}{2}\rho_{\vh{i}} b_{\vh{i}}^{(\vh{m})\,\dag}b_{\vh{i}}^{(\vh{n})}b_{\vh{i}}^{(\vh{k})\,\dag}b_{\vh{i}}^{(\vh{l})}\right).
\eea
\end{widetext}
In particular, for a homogeneous situation $\rho_i=\sigma$, the above master equation is a {\em nonlinear} equation for $\sigma$, since the Hamiltonian depends on the expectation values of the destruction operators in the various bands. We note that in contrast to the problem studied in \cite{Die10} the dissipative part is linear in $\rho$. This comes from neglecting scattering processes in which the atoms change lattice site and and the symmetry $\gamma_{\vh{i},\vh{j}}^{(\vh{k,l,m,n})}=\gamma_{\vh{j},\vh{i}}^{(\vh{m,n,k,l})}$. It is easy to show that the master equation \eqref{eq:mean_field_master_eq} preserves the trace and the mean particle number.

\begin{figure}[h]
  \centering
  \includegraphics[width=8cm]{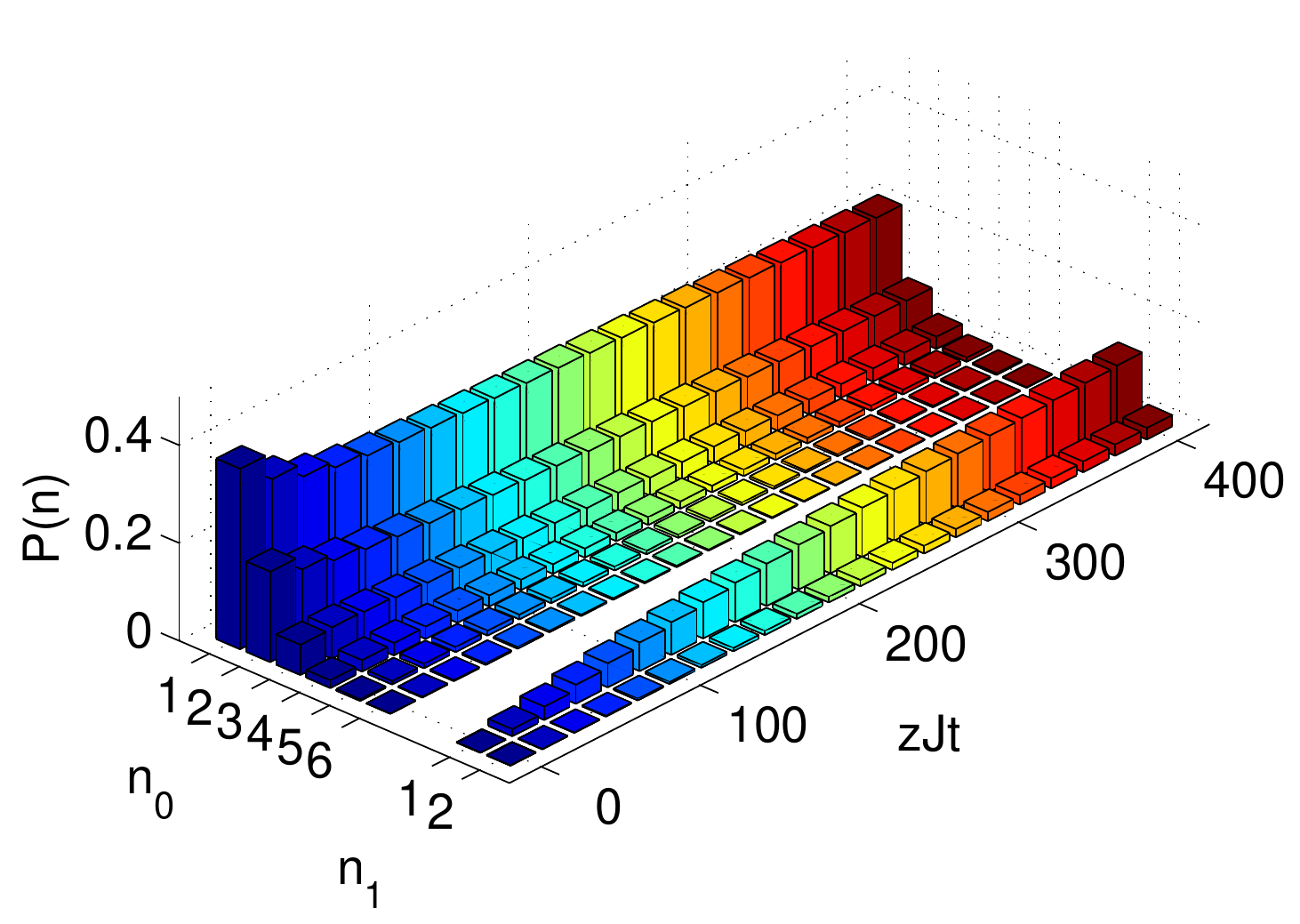}\caption{Particle number statistics for the lowest Band and one of the three first excited bands, computed for a noninteracting system as a function of time using a Gutzwiller ansatz for the system density operator. Initially the statistics in the lowest band are Poissonian, corresponding to a coherent state. The statistics in the lowest band change towards an exponential decay with $n_0$ as a function of time, while the population in the exited band increased. (Calculated with $\gamma^{(\vh{0,0,0,0})}=0.01zJ$ in a isotropic 3D lattice ($z=6$) of depth $V=10E_R$.)} \label{fig:gutznumstat0}
\end{figure}

\begin{figure}[h]
  \centering
  \includegraphics[width=8cm]{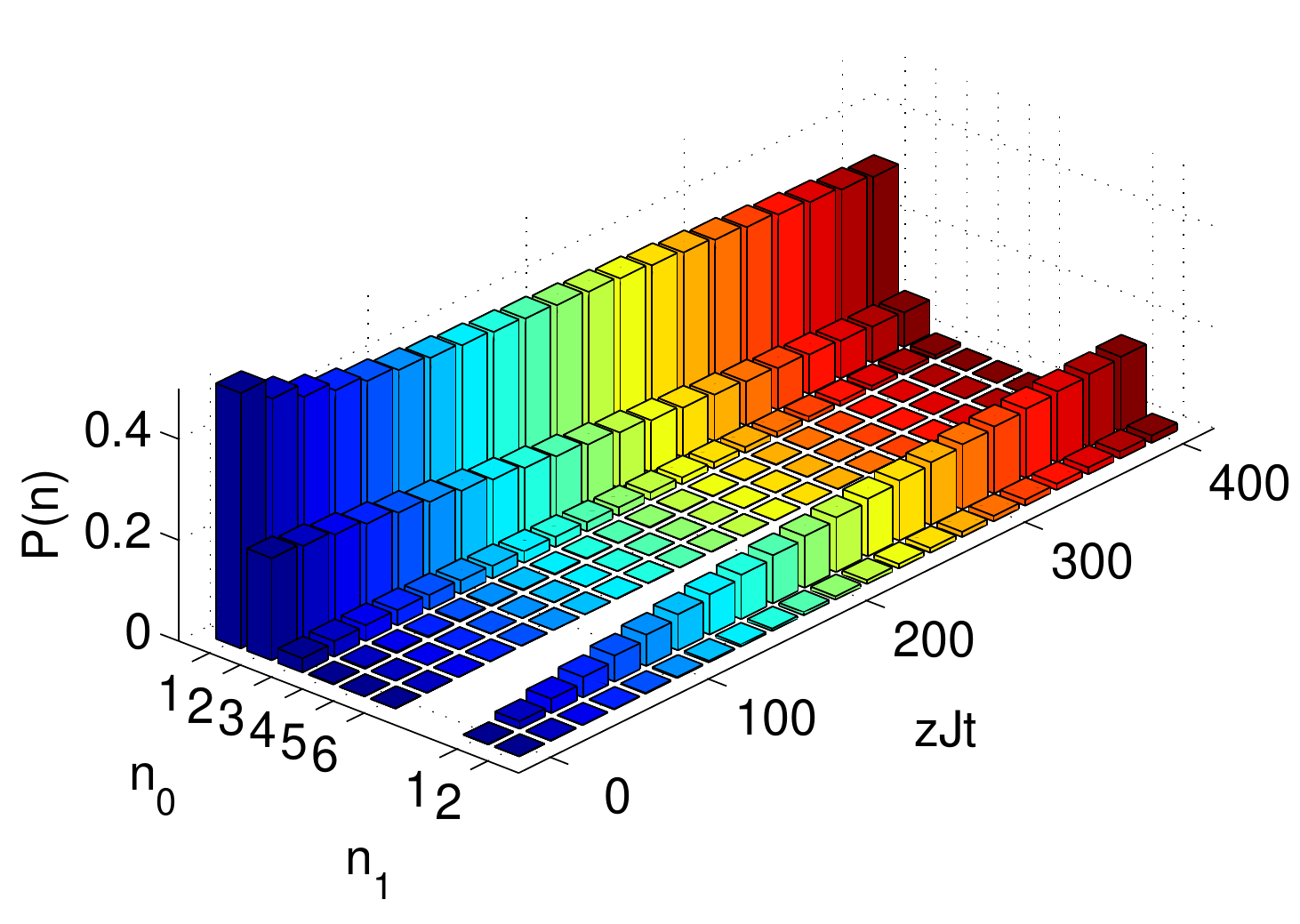}\caption{Particle number statistics for the lowest Band and one of the three first excited bands, computed for an interacting system with $U/(zJ)=1$ as a function of time using a Gutzwiller ansatz for the system density operator. As in Fig.~\ref{fig:gutznumstat0}, we choose $\gamma^{(\vh{0,0,0,0})}=0.01zJ$ in a isotropic 3D lattice ($z=6$) of depth $V=10E_R$.}\label{fig:gutznumstat1}
\end{figure}

\begin{figure}[h]
  \centering
  \includegraphics[width=8cm]{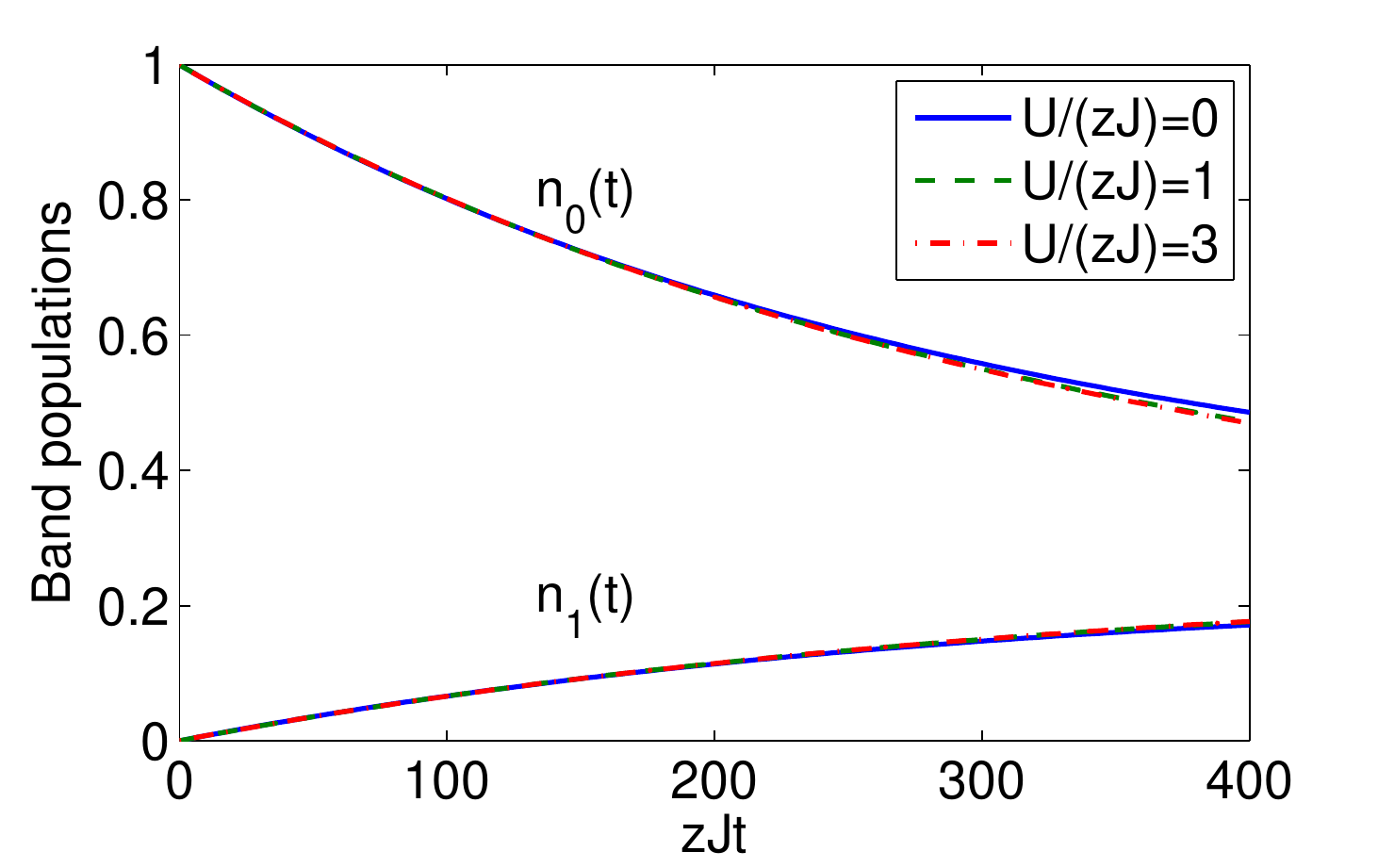}\caption{Time evolution of the mean number of particles per lattice site in the lowest and first excited bands, computed using a Gutzwiller ansatz for the system density operator, and plotted as a function of time for different interaction strengths. As the rate of population of higher bands is independent of the interaction strength, we see from the similarity of results for interacting and non-interacting systems that processes returning the atoms to the lowest band are very slow, leading to a lack of thermalization of energy transferred to the system by transferring particles to higher bands. As for Fig.~\ref{fig:gutznumstat0}, we choose $\gamma^{(\vh{0,0,0,0})}=0.01zJ$ in a isotropic 3D ($z=6$) lattice of depth $V=10E_R$.}\label{fig:gutzbandocc}
\end{figure}

\begin{figure}[h]
  \centering
  \includegraphics[width=8cm]{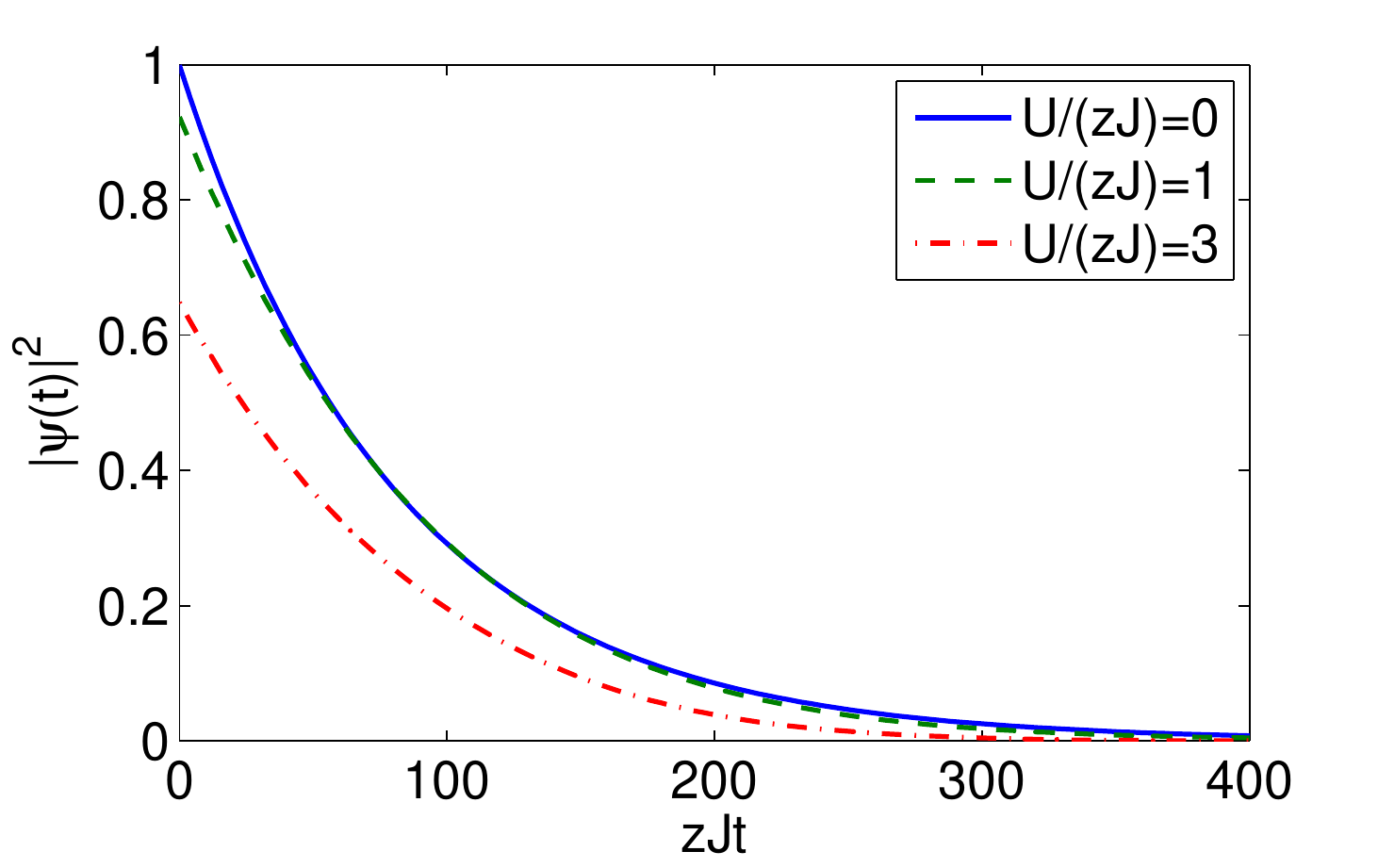}\caption{Evolution of the square modulus of the superfluid order parameter $|\psi(t)|^2$, computed using a Gutzwiller ansatz for the system density operator under the Eq.~(\ref{eq:mean_field_master_eq}) for different interaction parameters. The initial states are the corresponding Gutzwiller ground states. The different initial states have different non-vanishing superfluid order parameters, which in all cases decays on the timescale set by the localization rate in the lowest band. (Calculated with $\gamma^{(\vh{0,0,0,0})}=0.01zJ$ in a isotropic 3D lattice ($z=6$) of depth $V=10E_R$.)}
  \label{fig:gutzsuperfluidorder}
\end{figure}

\begin{figure}[h]
  \centering
  \includegraphics[width=8cm]{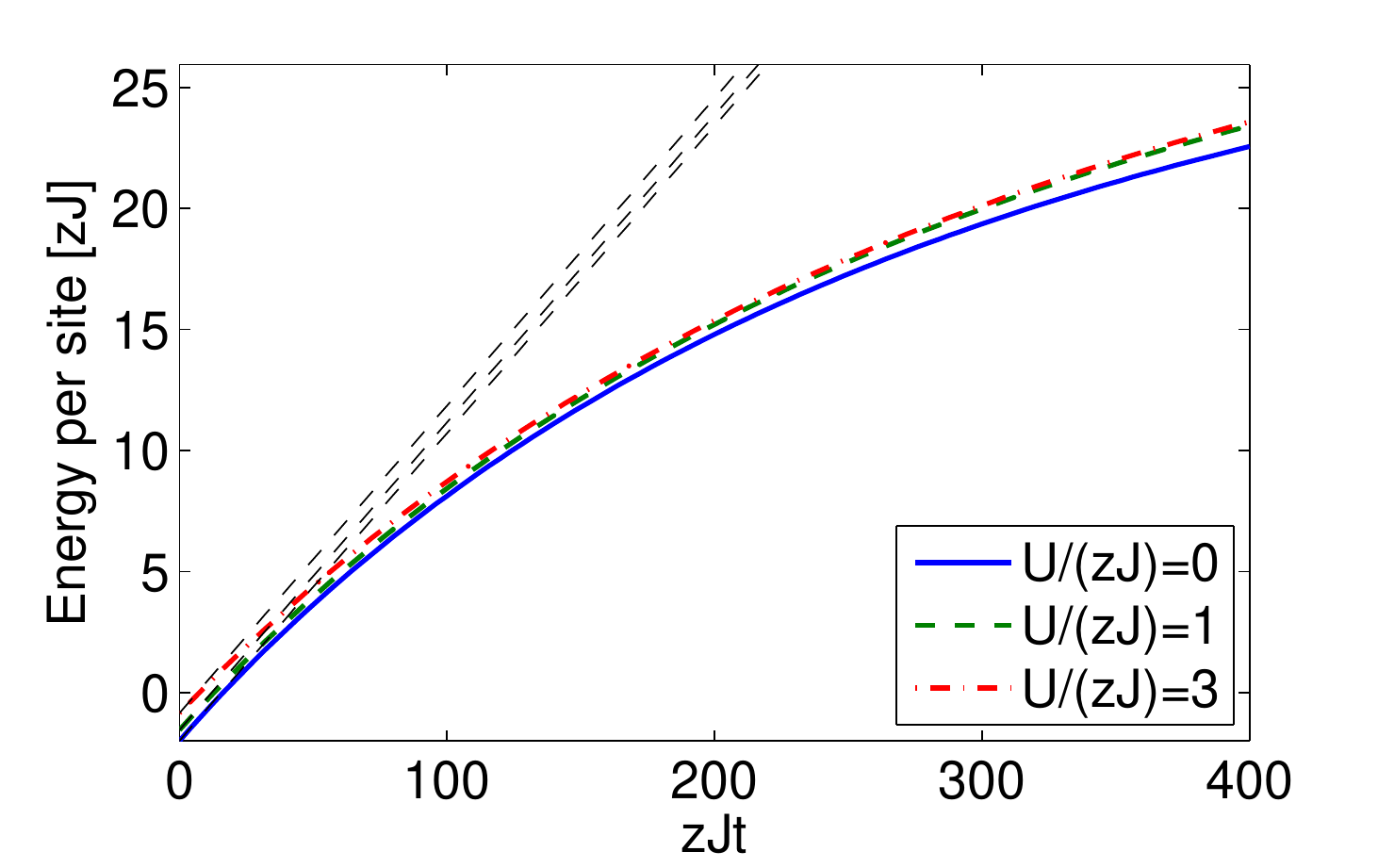}\caption{Evolution of the mean total energy per lattice site, computed using a Gutzwiller ansatz for the system density operator. The initial values depend on the interaction strength of the system. The increase in energy is the same in all cases, as we saw in Sec.~\ref{sec:total_heating}. Since we have included only one excited band, the total energy saturates, leading to the deviation from the linear increase of the exact solution. The thin dashed lines indicates the increase in energy as calculated in Sec.~\ref{sec:total_heating}. (Calculated with $\gamma^{(\vh{0,0,0,0})}=0.01zJ$ in a isotropic 3D lattice ($z=6$) of depth $V=10E_R$.) }\label{fig:gutzenergy}
\end{figure}

\begin{figure}[h]
  \centering
  \includegraphics[width=8cm]{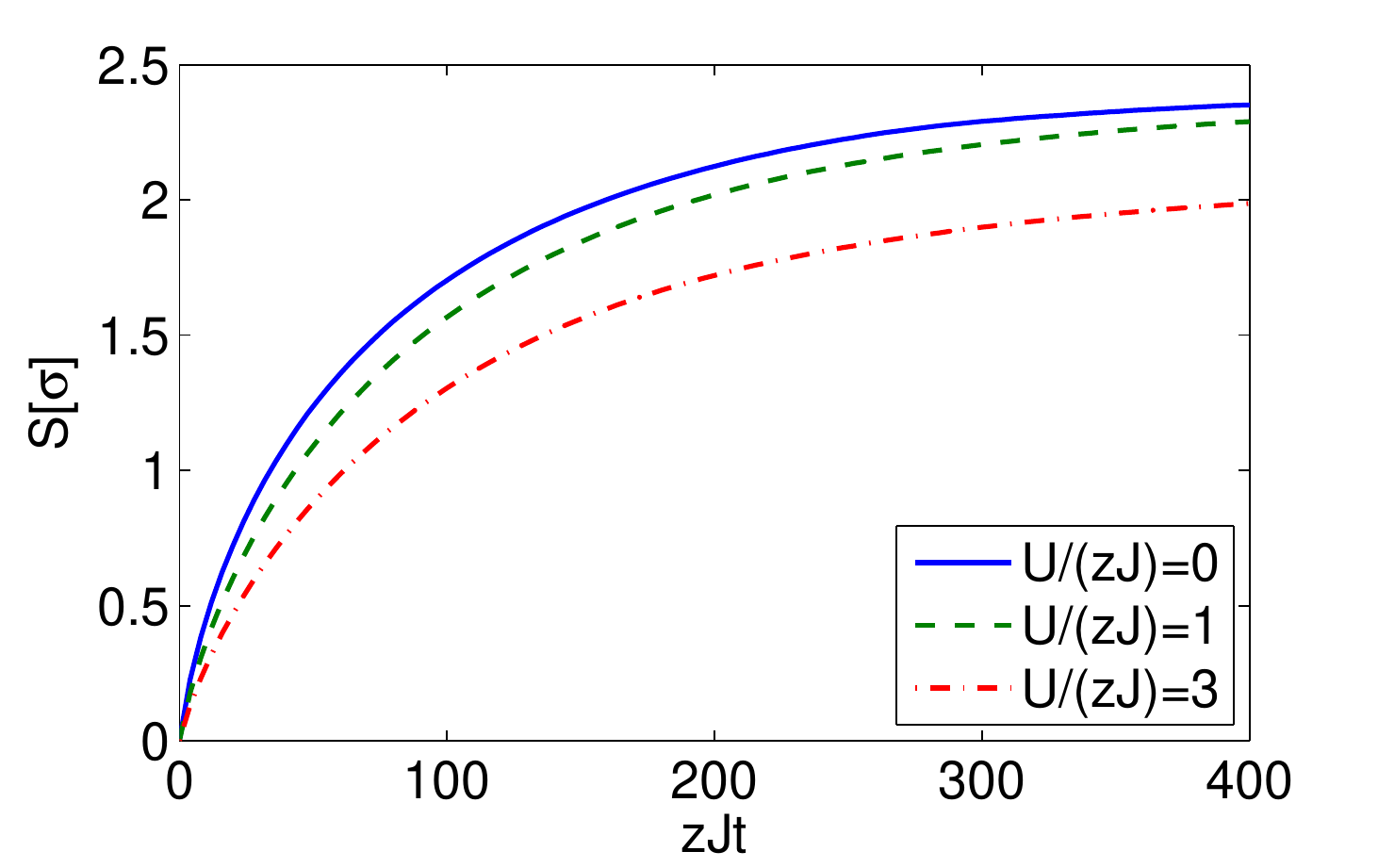}\caption{Increase of entropy per lattice site for different interactions, computed using a Gutzwiller ansatz for the system density operator. Starting in the ground state, the entropy is initially zero. Due to the different initial states the increase in entropy is higher for lower interaction parameters, reflecting the more significant change in more weakly interacting states due to spontaneous emission events. (Calculated with $\gamma^{(\vh{0,0,0,0})}=0.01zJ$ in a isotropic 3D lattice of depth $V=10E_R$.)}\label{fig:gutzentropy}
\end{figure}

Here we will consider the example of homogeneous system ($\rho_i\equiv\sigma$ for all sites) in an isotropic cubic 3D lattice ($z=6$), and will restrict our discussion to the lowest band and first three excited bands (which are degenerate for the isotropic case). In this way we include the dominant heating channels in our calculations, as in a relatively deep lattice the Lamb Dicke parameter $\eta$ is small, and the rates of heating to higher bands are significantly smaller, as discussed in Sec.~\ref{sec:singleparticle}. We begin each calculation with the Gutzwiller ground state at zero temperature for atoms in the lowest band of the lattice, with a given $U/J$ value at unit filling. The state is then propagated in time using the above method, with a Gutzwiller ansatz for the system density operator. The scattering rates are based on a red-detuned optical lattice with a depth of $V=10E_R$, setting $\gamma^{(\vh{0,0,0,0})}=0.01zJ$.

In Fig.~\ref{fig:gutznumstat0} and Fig.~\ref{fig:gutznumstat1} we show the evolution of the particle number distribution in the lowest and excited bands as a function of time for a non-interacting gas ($U=0$), and an interacting gas with $U/(zJ)=1$. For vanishing interaction $(U/(zJ)=0)$ the initial state exhibits Poissonian number statistics in the lowest band, $P(n_0)=\bra{n_0}\sigma\ket{n_0}=\mean{n_0}^{n_0} e^{-\mean{n_0}}/(n_0!)$. We start with one particle per lattice site in the lowest band so that $\mean{n_0}=1$. As the state evolves, we see that the number statistics change rapidly from Poissonian towards an exponential distribution. At the same time, the probability of population in the higher bands gradually increases. Similar behavior is observed when the interaction strength is increased to $U/(zJ)=1$. When we choose interaction strengths above the superfluid-Mott Insulator transition at unit filling, which is not shown here, the dynamics of the number statistics becomes trivial. In the Gutzwiller representation, we have exactly one particle per lattice site, and the only dynamics arising from spontaneous emission events is gradual population of excited bands.

In Fig.~\ref{fig:gutzbandocc} we show the time evolution of the mean particle number for the lowest and excited bands for different interaction strengths. We note that for different interaction strengths, these numbers are identical up to very long times. As the rate of population of higher bands is independent of the interaction strength, we see from the similarity of these results that that processes returning the atoms to the lowest band via collisions are very slow, leading to a lack of complete thermalization of the energy transferred to the system, as was discussed in Sec.~\ref{sec:lack_of_thermalization}.

In Figs.~\ref{fig:gutzsuperfluidorder}--\ref{fig:gutzentropy} we quantify the heating for states of different initial $U/(zJ)$ values via different quantities. The quantity corresponding to off diagonal long range order is $\sum_{\vh{n}}\mean{b^{(\vh{n})}}\equiv\psi$ and we identify this quantity as an order parameter for superfluidity.
In Fig.~\ref{fig:gutzsuperfluidorder} we show the evolution of this superfluid order parameter, where we clearly see the destruction of superfluidity for atoms in the lowest band as a result of the heating processes. This is analogous to the destruction of long-range order discussed in \ref{sec:quantspdm}. In the Mott Insulator phase, the superfluid order parameter is zero at the beginning, and remains zero throughout the evolution. In Fig.~\ref{fig:gutzenergy}, we show the total energy increase in the system over time. These results agree well for short times with the results obtained in Sec.~\ref{sec:total_heating}. A nice feature of the Gutzwiller ansatz is that it is simple also to calculate the increase in entropy for the system, as we start from a pure state and heat the system into a mixed state. The entropy per lattice site is plotted in Fig.~\ref{fig:gutzentropy}, and describes the same basic behavior as Figs.~\ref{fig:gutzsuperfluidorder} and \ref{fig:gutzenergy}.

In summary, we find that a product ansatz for the system density operator in the spirit of a Gutzwiller meanfield treatment gives a simple semi-quantitative picture for the heating process. Including higher Bloch bands, we observe quantitatively that particles heated to higher bands are not transferred back to the lowest band on typical experimental timescales, even in the presence of significant interactions. As in the previous section, we see that scattering in the lowest band gives rise to a destruction of superfluidity in the system, here characterized via the superfluid order parameter. This is always zero in the Mott Insulator phase, and the key properties of this phase in the Gutzwiller description change only in that particles can be heated to higher bands. In this sense, we see that more strongly interacting states are significantly more robust against heating due to spontaneous emission events.

\section{Summary and Outlook}
\label{sec:conclusion}

We have shown that the heating of atoms in optical lattices due to spontaneous emission events is strongly dependent on the characteristics of the lattice, especially the detuning of the lattice beams, and also on the many-body characteristics of the state. Because atoms scattered to higher Bloch bands will not thermalize with atoms remaining in the lowest band on experimental timescales, it is not sufficient to compute the total rate of energy increase in order to determine the change in the many-body state. Instead, the heating can be characterized, e.g., by computing characteristic correlation functions, such as the single particle density matrix for bosons.

We found that the higher scattering rate for red-detuned lattices as opposed to blue-detuned optical lattices corresponds to a much more rapid breakdown in off-diagonal order in a superfluid state, due to the localizing effect of spontaneous emission events. In contrast, a Mott Insulator state, where the atoms are already exponentially localized, can be strongly robust against spontaneous emission events.

In an experiment, other design considerations will have to be taken into account when choosing the lattice detuning, e.g., the different rates of light-assisted collisions $\gamma_2$ for red- and blue-detuned laser light. This interplay is particularly interesting because light-assisted collisions tend to be more prominent for blue-detuned light, which is where the rate of spontaneous emissions is lowest. For production of states where the atoms are exponentially localized at different lattice sites, red detuned lasers could be used without strong adverse effects. However, for production of states with off-diagonal long-range order, we have shown here that the laser detuning is an important consideration. In the future, the quantum trajectories methods we have used here could be extended to include two-body loss terms, and used to analyze the competition between different heating mechanisms in these experiments.

Another key future direction will be the investigation of heating of fermionic species. The results here give an indication that states in which atoms are localized, e.g., a Mott Insulator state with possible additional spin-ordering could, under favorable circumstances, be relatively robust against spontaneous emission events.

\label{txt:outlook}

\section*{Acknowledgments}
We thank Matthias Troyer, Lode Pollet, and the groups of Tilman Esslinger and Immanuel Bloch for helpful and motivating discussions. This work was supported by the Austrian Science Fund through SFB F40 FOQUS and EUROQUAM\_DQS (I118-N16), and by a grant from the US Army Research Office with funding from the DARPA OLE program.

\begin{appendix}

\section{Calculation of the total heating rate}\label{ap:heting_rate}
The total heating rate (\ref{eq:energy_increase_result}) is calculated as the change rate of $H$, the hermitian part of $H_{\rm eff}$ in Eq.~\eqref{eq:master_equation}:
\begin{align}\label{eq:heating_calc_1}
&\frac{d}{dt}\langle \hat H \rangle=
\mathrm{Tr}\{\hat H \mathcal{L}_1\rho\}=\no\\
&=-\frac{1}{2}\frac{\Gamma}{4\Delta^2}\mathrm{Tr}\{
\iint d^3xd^3yF(k(\vh{x}-\vh{y}))
\Omega(\vh{x})\Omega(\vh{y})\no\\
&\qquad\times\left[\left[\hat H ,\hat{\psi}^{\dag}(\vh{x})\hat{\psi}(\vh{x})\right],
\hat{\psi}^{\dag}(\vh{y})\hat{\psi}(\vh{y})\right]\rho\}.
\end{align}
 We use the approximation $k_{eg}\approx k_L\equiv k$ and introduce the notation $\mathcal{F}(\vh{x},\vh{y})\equiv F(k(\vh{x}-\vh{y}))\Omega(\vh{x})\Omega(\vh{y})$. Noting that $F(0)=1; \nabla F|_0=0$ and $\triangle F|_0=-k^2$ and using Maxwell's equation $\triangle \Omega(\vh{x})=-k^2\Omega(\vh{x})$ we readily derive the following relations:
 \begin{align}\label{eq:heating_relation_1}
\triangle_x\mathcal{F}(\vh{x},\vh{y})|_{\vh{y}=\vh{x}}&=-2k^2\Omega(\vh{x})^2,\\\label{eq:heating_relation_2}
\nabla_x\mathcal{F}(\vh{x},\vh{y})|_{\vh{y}=\vh{x}}&=\Omega(\vh{x})\nabla_x\Omega(\vh{x}).
\end{align}
Here the $\triangle$ denotes the Laplacian.
Further we have:
\begin{align}
&[\hat H,\hat{\psi}^{\dag}(\vh{x})\hat{\psi}(\vh{x})]=\nonumber\\
&=\frac{-1}{2m}\left(\left(\triangle_{\vh{x}}\hat{\psi}^{\dag}(\vh{x})\right)\hat{\psi}(\vh{x})
-\hat{\psi}^{\dag}(\vh{x})\left(\triangle_{\vh{x}}\hat{\psi}(\vh{x})\right)\right).
\end{align}
Using this together with the relations (\ref{eq:heating_relation_1}) and (\ref{eq:heating_relation_2}) in Eq.~(\ref{eq:heating_calc_1})
we find after partial integration:
\begin{align}
&\frac{d}{dt}\langle \hat H \rangle=\frac{1}{2}\frac{\Gamma}{2m\Delta^2}\mathrm{Tr}\{\int d^3x\left(2k^2\Omega(\vh{x})^2\hat{\psi}^{\dag}(\vh{x})\hat{\psi}(\vh{x})
+\right.\no\\
&\left.+(\nabla_x\Omega(\vh{x})\cdot\nabla_x\Omega(\vh{x})+
\Omega(\vh{x})\triangle_x\Omega(\vh{x}))\hat{\psi}^{\dag}(\vh{x})\hat{\psi}(\vh{x})\right)\rho\}.\\
\end{align}
In the lattice we have the relation $2k^2\Omega(\vh{x})^2+\nabla_x\Omega(\vh{x})\cdot\nabla_x\Omega(\vh{x})+
\Omega(\vh{x})\triangle_x\Omega(\vh{x})= k^2$, which leads to:
\begin{align}
\frac{d}{dt}\langle \hat H \rangle&=\frac{1}{2}\frac{\Gamma\Omega_0^2}{4\Delta^2 2m}\int d^3x2k^2\mathrm{Tr}\{\hat{\psi}^{\dag}(\vh{x})\hat{\psi}(\vh{x})\rho\}=\no\\
&=\frac{\Gamma\Omega_0^2}{4\Delta^2}\frac{k^2}{2m}N.
\end{align}
This is the result stated in Eq.~(\ref{eq:energy_increase_result}).

\end{appendix}

\end{document}